\documentclass[a4paper,conference]{IEEEtran}

\usepackage[american]{babel}
\usepackage{hyphenat}

\usepackage{tikz}
\usetikzlibrary{positioning}

\usepackage{algorithmic}
\usepackage{algorithm}
\usepackage{array}

\usepackage{subfigure}

\usepackage{textcomp}
\usepackage{url}
\usepackage{verbatim}
\usepackage{graphicx}

\usepackage{cite}
\usepackage[utf8]{inputenc}
\usepackage[T1]{fontenc}
\usepackage{textcomp}
\usepackage{microtype}
\usepackage{calc}
\usepackage[normalem]{ulem}

\usepackage{amsmath}
\usepackage{amssymb}
\usepackage{amsfonts}
\usepackage{amsthm}

\usepackage{mathtools}

\usepackage[capitalize,noabbrev,nameinlink]{cleveref}

\RequirePackage{xstring}
\RequirePackage{xparse}
\RequirePackage[]{acro}
\makeatletter
\@ifpackagelater{acro}{2019/02/17}{
	\NewDocumentCommand\acrodef{mO{#1}mG{}}{\DeclareAcronym{#1}{short={#2}, long={#3}, foreign-plural={}, #4}}
}{
	\NewDocumentCommand\acrodef{mO{#1}mG{}}{\DeclareAcronym{#1}{short={#2}, long={#3}, #4}}
}
\makeatother
\acrodef{AIFS}{Arbitration Inter-Frame Spacing}
\acrodef{AoI}{Age of Information}
\acrodef{AWGN}{Additive White Gaussian Noise}
\acrodef{BS}{Base Station}
\acrodef{BSS}{Basic Service Set}
\acrodef{CAM}{Cooperative Awareness packet}
\acrodef{CBF}{Contention-Based Forwarding}
\acrodef{CBR}{Channel Busy Ratio}
\acrodef{CCDF}{Complementary Cumulative Distribution Function}
\acrodef{CCH}{Control Channel}
\acrodef{CDMA}{Code Division Multiple Access}
\acrodef{CDF}{Cumulative Distribution Function}
\acrodef{C-ITS}{Cooperative Intelligent Transportation System}{short-plural-form={C-ITS}}
\acrodef{COV}{Coefficient Of Variation}
\acrodef{CPM}{Collective Perception packet}
\acrodef{CPU}{Central Processing Unit}
\acrodef{DSRC}{Dedicated Short-Range Communication}
\acrodef{DTMC}{Discrete Time Markov Chain}
\acrodef{FCFS}{First Come First Served}
\acrodef{GFMA}{Grant Free Multiple Access}
\acrodef{GNCU}{Google Normalized Computing Unit}
\acrodef{GNMU}{Google Normalized Memory Unit}
\acrodef{i.i.d.}{independent and identically distributed}
\acrodef{IAT}{Inter arrival times}
\acrodef{JIQ}{Join Idle Queue}
\acrodef{JSQ}{Join Shortest Queue}
\acrodef{LCFS}{Last Come First Served}
\acrodef{LCFSwO}{Last Come First Served with Overwrite}
\acrodef{LDM}{Local Dynamic Map}
\acrodef{LWL}{Least Work Left}
\acrodef{mMTC}{Massive machine-type Communication}
\acrodef{M2M}{Machine-to-Machine}
\acrodef{MAC}{Medium Access Control}
\acrodef{MCM}{Maneuver Coordination packet}
\acrodef{MCS}{Modulation and Coding Scheme}
\acrodef{MPR}{Multi-Packet Reception}
\acrodef{MRT}{Mean Response Time}
\acrodef{MUSA}{Multi User Shared Access}
\acrodef{NOMA}{Non-Orthogonal Multiple Access}
\acrodef{NoB}{No Buffering}
\acrodef{OBU}{On-Board Unit}
\acrodef{OMA}{Orthogonal Multiple Access}
\acrodef{PDF}{Probability Density Function}
\acrodef{PDMA}{Pattern Division Multiple Access}
\acrodef{PER}{Packet Error Rate}
\acrodef{PSJF}{Preemptive-Shortest-Job-First}
\acrodef{RAM}{Random Access Memory}
\acrodef{RR}{Round Robin}
\acrodef{RSU}{Road Side Unit}
\acrodef{SCH}{Service Channel}
\acrodef{SCOV}{Squared Coefficient Of Variation}
\acrodef{SA}{Slotted ALOHA}
\acrodef{SD}{Slowdown}
\acrodef{SIC}{Successive Interference Cancellation}
\acrodef{SITA}{Size Interval Task Assignment}
\acrodef{SNR}{Signal to Noise Ratio}
\acrodef{SNIR}{Signal to Noise plus Interference Ratio}
\acrodef{SRPT}{Shortest Remaining Processing Time}
\acrodef{TRG}{Two-Ray Ground}
\acrodef{V2I}{Vehicle-to-Infrastructure}
\acrodef{V2V}{Vehicle-to-Vehicle}
\acrodef{V2X}{Vehicle-to-Everything}
\acrodef{VANET}{Vehicular Ad Hoc Network}
\acrodef{VLC}{Visible Light Communication}
\acrodef{UAV}{Unmanned Aerial Vehicle}

\def\todoCtd#1{%
	TODO: #1%
	\ifx&#1&.\fi%
	\endgroup
	\cbend
	\relax
}

\begin{document}

\title{Dispatching Odyssey: Exploring Performance in Computing Clusters under Real-world Workloads}

\author{%
\IEEEauthorblockN{%
	Mert Yildiz\IEEEauthorrefmark{1}%
	,
	Alexey Rolich\IEEEauthorrefmark{1}%
	,
	Andrea Baiocchi\IEEEauthorrefmark{1}%
}%

\IEEEauthorblockA{
	\IEEEauthorrefmark{1}\small{Dept.\ of Information Engineering, Electronics and Telecommunications (DIET), University of Rome Sapienza, Italy}
}%


%
\texttt{%
	\{mert.yildiz,alexey.rolich,andrea.baiocchi\}%
	@uniroma1.it%
}%
}

\IEEEpubid{\makebox[\columnwidth]{978-3-948377-03-8/19/\$31.00 \copyright 2025 ITC \hfill} \hspace{\columnsep}\makebox[\columnwidth]{ }}

\maketitle

\begin{tikzpicture}[overlay, remember picture]
  \node[anchor=south] (botAnchor) at (current page.south) {};
  \node[above=of botAnchor, node distance=1cm] {%
    \footnotesize 978-3-948377-03-8/19/\$31.00 \copyright
    ~2025 ITC
  };
\end{tikzpicture}

\begin{abstract}\nohyphens{%
Recent workload measurements in Google data centers provide an opportunity to challenge existing models and, more broadly, to enhance the understanding of dispatching policies in computing clusters. Through extensive data-driven simulations, we aim to highlight the key features of workload traffic traces that influence response time performance under simple yet representative dispatching policies. For a given computational power budget, we vary the cluster size, i.e., the number of available servers. A job-level analysis reveals that Join Idle Queue (JIQ) and Least Work Left (LWL) exhibit an optimal working point for a fixed utilization coefficient as the number of servers is varied, whereas Round Robin (RR) demonstrates monotonously worsening performance. Additionally, we explore the accuracy of simple G/G queue approximations. When decomposing jobs into tasks, interesting results emerge; notably, the simpler, non-size-based policy JIQ appears to outperform the more “powerful” size-based LWL policy. Complementing these findings, we present preliminary results on a two-stage scheduling approach that partitions tasks based on service thresholds, illustrating that modest architectural modifications can further enhance performance under realistic workload conditions. We provide insights into these results and suggest promising directions for fully explaining the observed phenomena.
}\end{abstract}


\begin{tikzpicture}[overlay, remember picture]
\path (current page.north) node (anchor) {};
\node [below=of anchor] {2025 36th International Teletraffic Congress (ITC 36)};
\end{tikzpicture}

%

\section{Introduction}
\label{sec:introduction}

Finding effective and easily implementable algorithms for distributing the workload in a data center is among the top issues in the framework of the evolution of cloud and edge computing.
Given a cluster of servers to which a sequence of jobs is offered, dispatching policies take care of assigning each job (possibly broken down into several tasks) to one server.
Scheduling policies specify how each server deals with its own assigned jobs/tasks.

While many dispatching and scheduling policies have been defined and analyzed under different assumptions (e.g., see \cite{HarcholBalter2013,Zhou2017,KUMAR2019,Tang2024}), still many open issues are on the table \cite{HarcholBalter2021}.
A thorough understanding of multi-server cluster systems remains incomplete due to the difficulty of modeling the essential characteristics of workloads, particularly when they involve complex policies.
The viability of model analysis often requires strongly simplifying assumptions \cite{Hyytia2020,Choudhury2021,Bilenne2021,AbdulJaleel2022}, e.g., Poisson arrivals or negative exponential service times.
Assessing whether those assumptions match, at least qualitatively, real system performance is an essential step of the design and optimization of computing clusters.
To this end, the availability of high-quality massive workload measurements is valuable for characterizing real-world workloads in data centers.

A detailed analysis of the Google Borg dataset in \cite{wilkes2019clusterdata, Tirmazi2020, Sfakianakis2021Tracebased, yildiz2024} reveals discrepancies between basic queuing model assumptions and the characteristics of real-world data. 
Specifically, job arrivals do not follow a Poisson process, and the service time interpreted as CPU requirements in \cite{HarcholBalter2021,yildiz2024}, does not adhere to a negative exponential distribution. 
On the contrary, the distribution of service times is highly variable and exhibits an extremely heavy-tailed behavior, whereby a small fraction of the very largest jobs comprises most of the load. 
In prior empirical studies of compute consumption and file sizes \cite{crovella1998,harcholbalter1996,harcholbalter1996Downey,harcholbalter1999,harcholbalter2003}, the authors found that the top 1\% of jobs comprise 50\% of the load.
The heavy-tailed property exhibited in Google data centers today is even more extreme than what was seen in \cite{crovella1998,harcholbalter1996,harcholbalter1996Downey,harcholbalter1999,harcholbalter2003}. 
For Google jobs today, the largest (most compute-intensive) 1\% of jobs comprise about 99\% of the compute load \cite{Tirmazi2020}, which might cause a bigger discrepancy between the prediction of analytical models and data-driven simulations.

In this paper, we consider real workload measurements and data-driven simulations to evaluate the performance of a server system comprising a dispatcher and a cluster of homogeneous servers.
The focus is on comparing three representative dispatching policies, of different complexity: size-based (\ac{LWL}), non-size-based, but stateful (\ac{JIQ}), and non-size-based stateless (\ac{RR}).
Our analysis reveals that dispatching has a significant impact on performance, highlighting the need for dedicated attention to this aspect of system design.

The main research questions that we try to answer in this work are as follows. Do simple analytical models provide insight consistent with what real-world workloads reveal through data-driven simulations? What are the main factors affecting response time performance? Based on the understanding gained from the first two points, can we improve performance through system architecture design, even using very simple policies?

With reference to this last question, as noted in \cite{HarcholBalter2022}, one of the most common queuing theory questions asked by computer systems practitioners is "How can I favor short jobs?".
This target can be tackled by using either \ac{SRPT} or \ac{PSJF} scheduling policies; along with \ac{LWL} or \ac{SITA} dispatching policies. 
However, in practical scenarios, when jobs arrive at computing clusters, their sizes are typically unknown to the system, rendering these solutions impractical in real-world settings.
For these reasons, \ac{FCFS} is widely used in many organizations, including Google, where the Borg data center scheduler operates a large central \ac{FCFS} queue \cite{HarcholBalter2021}.
However, the simple and easily implementable \ac{FCFS} scheduling policy suffers from service time variability.
As a solution, we propose a novel two-stage dispatching and scheduling approach, stemming from the insight gained from the analysis presented in this work.
This two-stage multi-server system combines the simplicity of the \ac{RR} dispatching policy with \ac{FCFS} scheduling, where neither the dispatcher nor the scheduler needs prior knowledge of job characteristics. 
Yet, it is shown that it outperforms single-state server clusters, even using more sophisticated dispatching policies.


To gain insight and motivate our proposed two-stage architecture, first, we focus on job-level performance analysis.
\ac{LWL}, \ac{JIQ} and \ac{RR} dispatching policies are compared for a single-stage cluster of $n$ servers, as the number $n$ varies, under the constraint that the average load per server be the same (i.e., the overall computational power of the cluster is fixed).
The obtained results provide insight into the effect of parallelism on response time under real-world workloads.
This analysis also helps us identify the root causes of discrepancies between simple analytical models and simulation results.
It appears that service time process structure is the main source of discrepancy, while inter-arrival times of jobs could be replaced with a Poisson stream of arrivals with little impact on the accuracy of predictions of the analytical model.

In the second step, we account for the internal structure of jobs, which are decomposed into independent tasks.
The most significant finding is that, when accounting for task level, \ac{JIQ} outperforms \ac{LWL} in spite of being heavy-traffic delay sub-optimal.
Even more strikingly, when introducing multi-stage architecture, even \ac{RR} outperforms \ac{LWL} in a substantial way, by properly optimizing the meta-parameters of the architecture.
This points to a promising direction for developing high-performance systems, namely smart design of the architecture, rather than smart (possibly complex) policies.

The rest of the paper is organized as follows.
\cref{sec:dataset} provides a brief explanation of the main features of workload measurement data.
\cref{sec:modeldescription} gives an account of the considered model setting.
Numerical results of data-driven simulations and their comparison with the model are presented in \cref{sec:numres}.
Finally, conclusions are drawn in \cref{sec:conclusions}.

%

\section{Workload dataset}
\label{sec:dataset}

This section describes the Google traffic dataset released in 2020, capturing one month of user and developer activity in May 2019. 
It provides detailed resource usage data, such as \ac{CPU} and \ac{RAM}, across eight global data centers ("Borg cells").

Traffic data includes five tables detailing users' resource requests, machine processing, and task evolution within Borg's scheduler. 
Users submit jobs comprising one or more tasks (instances), each requiring specific \ac{CPU} time and memory space. 
Resource units are based on \ac{GNCU} (default machine computational power) and \ac{GNMU} (memory units).
 \ac{CPU} time associated with the execution of a task assumes that servers are equipped with one \ac{GNCU}.
 
Jobs follow a life cycle: tasks are queued or processed based on load and classified upon completion as ``FINISH'' (success) or ``KILL,'' ``LOST,'' or ``FAIL'' (failure) \cite{wilkes2019clusterdata}. Only ``FINISH'' tasks were analyzed, as they provide reliable resource data. The dataset includes task identifiers, \ac{CPU} time, assigned memory (min, max, avg over 300-sec windows), and 1-sec samples of resource usage.

Digging into Google's data, the workloads have been reconstructed with the following key features:\footnote{The dataset for all eight data centers is available at \textcolor{blue}{\url{https://github.com/MertYILDIZ19/Google_cluster_usage_traces_v3_all_cells}}.}
\begin{itemize}
	\item $J_j$ - number of tasks belonging to job $j$.
	\item $A_{ij}$ - Arrival time (in seconds) of task $i$ of job $j$.
	\item $Z_{ij}$ - \ac{CPU} time (in seconds) required to process task $i$ of job $j$ on a reference \ac{CPU} equipped with one \ac{GNCU}.
\	\item $M_{ij}$ - maximum amount of memory space required by task $i$ of job $j$, expressed in \acp{GNMU}.
\end{itemize}

This study focuses on simplified workload description by considering only arrival times and \ac{CPU} requirements. 
Memory-based multi-dimensional models are deferred to future research. 
Specifically, we analyze workload both at the job level, disregarding the breakdown of each job into separate tasks, and at the task level, where we allow tasks belonging to the same job to be assigned to different servers.
The characterization of the workload at the job level consists of job arrival time and job \ac{CPU} time.
The former is well defined since measured data provide evidence that arrival times $A_{ij}$ depend only on $j$, i.e., all tasks belonging to the same job arrive simultaneously (within the accuracy of timestamps).
As for job \ac{CPU} time, it is simply defined as the sum of \ac{CPU} times required by all tasks belonging to the job.

The whole considered dataset (31 days worth of measured activity in Borg cell c) consists of 4399670 jobs, comprising 7010742 tasks.
Most jobs consist of a single task (96.6\%), yet ``monster'' jobs are also recorded, with the largest ones comprising up to 11160 tasks.
The mean job \ac{CPU} time is 10.83 s.
Only 13.25\% of jobs require \ac{CPU} time larger than the mean, while 99\% of jobs have \ac{CPU} time less than 32.26 s, 99.9\% of jobs less than 452.95 s, and 99.99\% of jobs less than 4974.2 s.
The largest 0.1\% of jobs is responsible for 67.6\% of the overall required \ac{CPU} effort.
These few numbers give evidence of the extreme variability of workload described in Google's data.


%

\section{Model description}
\label{sec:modeldescription}

In this section, we provide simple analytical formulas to evaluate the mean response time of single-stage server clusters.
A flow of jobs is submitted to a cluster of $n$ fully accessible and identical servers.
New jobs arrive with mean rate $\lambda$.
Let also $\sigma_A$ be the standard deviation of inter-arrival times and $C_A = \lambda \sigma_A$ the corresponding \ac{COV}.

A job comprises $J \ge 1$ tasks.
Task $i$ belonging to a given job is characterized by the time $Z_i$ required to be processed on a reference processing core having a capacity of 1 \ac{GNCU}.
The service time of a job on a server of capacity $\mu$ \ac{GNCU} is given by
\begin{equation}
X = \sum_{ i = 1 }^{ J }{ \frac{ Z_i }{ \mu } } = \frac{ Y }{ \mu }
\end{equation}
where $Y = \sum_{ i = 1 }^{ J }{ Z_i  }$ is the service time of the job on a reference server with a capacity of 1 \ac{GNCU} (i.e., $\mu = 1$).

We assume that $Y$ can be characterized as a continuous random variable with \ac{CDF} $F_Y(t) = \mathcal{P}( Y \le t )$, \ac{CCDF} $G_Y(t) = \mathcal{P}( Y > t )$ and \ac{PDF} $f_Y(t)$.
The mean of $Y$ is denoted with $\mathrm{E}[ Y ]$ and standard deviation given by $\sigma_Y$.
We define also the \ac{COV} of $Y$ as $C_Y = \sigma_Y / \mathrm{E}[ Y ]$.
Note that $G_X(t) = G_Y( \mu t )$, where $X$ is job service time on a generic server having a capacity of $\mu$ \ac{GNCU}.

It is assumed that all servers use \ac{FCFS} policy and have the same capacity of $\mu$ \acp{GNCU}.
The parameters $\mu$ and the number $n$ of servers are set so that
\begin{equation}
\label{eq:overallcomppowerbudget}
\frac{ \lambda \mathrm{E}[ Y ] }{ n \, \mu } = \rho_0
\end{equation}
for a given fixed value of $\rho_0$.
Note that, for any considered server arrangement, the overall service capacity $n \mu$ is a constant, given the user demand characteristics $\lambda$ and $\mathrm{E}[Y]$, and the target utilization coefficient $\rho_0$.

Since $X = Y/\mu$ is the service time of a job, from \cref{eq:overallcomppowerbudget} it follows that the mean service time is given by
\begin{equation}
\label{eq:meanservicetimeX}
\mathrm{E}[ X ] = \frac{ \mathrm{E}[ Y ] }{ \mu } = \frac{ n \, \rho_0 }{ \lambda }
\end{equation}

Finally, since $\mu$ is a constant for a given number of servers, the \ac{COV} of the service time $X$ is the same as that of $Y$, $C_X = C_Y$.
Summing up, the first two moments of inter-arrival times and service times of a workload at the job level are given by $1/\lambda$, $C_A$, $\mathrm{E}[Y]$, and $C_Y$.

We define the job response time $R$ as the time elapsing since job arrival until the job is fully completed (i.e., all of its tasks are finished).
We do not consider the possibility that the job is killed before being completed.

The analytical expression given in the following is derived following Marchal's approximation of G/G queuing systems for mean response time \cite{Marchal1985}.
The essential idea is to start from the expression of the mean response time.

It is therefore assumed that arrivals and service times form \emph{renewal} sequences, disregarding any possible correlation structure of the arrival and service process.
Moreover, it is assumed that inter-arrival times are independent of service times.

All models considered in the following deal with a job as a whole, i.e., all tasks belonging to that job are assigned to the same server and run one after the other (server use \ac{FCFS} policy).
While not exploiting the break-up of a job into smaller tasks, which enhances flexibility, this analysis is easier to interpret and gain insight into the relationship of performance predicted by analytical models and real-world measurements.

\subsection{Round Robin (RR) dispatching}
\label{subsec:singlestageRR}

In this case, the arrival stream of the job is distributed across $n$ servers according to a \ac{RR} dispatching policy.
The $j$-th arriving job is assigned to server $1+( j \mod n )$.
Hence, the mean arrival rate at each server is $\lambda_1 = \lambda / n$ and the \ac{COV} of arrivals to a server is given by $C_{A_1} = C_A/\sqrt{n}$.
The mean service time is $\mathrm{E}[ X_1 ] = \mathrm{E}[ Y ] / \mu = n \, \rho_0 / \lambda$.
The service time \ac{COV} is $C_{X_1} = C_X = C_Y$.
The server utilization coefficient is $\rho_1 = \lambda_1 \mathrm{E}[ X_1 ] = \rho_0$, same for all servers.
Since dispatching is instantaneous, the mean response time reduces to the time through a server.
It is given by
\begin{equation}
\mathrm{E}[ R_{\text{RR}} ] = \mathrm{E}[ X_1 ] + \mathrm{E}[X_1] \, \frac{ \rho_1 }{ 1 - \rho_1 } \phi_1
\end{equation}
where $\phi_1$ is the scaling factor of the mean waiting time to account for variability of inter-arrival and service times:
\begin{equation}
\phi_1 = \frac{ ( 1 + C_{X_1}^2 ) ( C_{A_1}^2 + \rho_1^2 C_{X_1}^2 ) }{ 2 ( C_{A_1}^2 + \rho_1^2 C_{X_1}^2 ) }
\end{equation}

This result can be expressed in terms of the parameters of the input workload, $\lambda$, $C_A$, $\mathrm{E}[Y]$, $C_Y$ and the target $\rho_0$, which is a design parameter.
The final expression is
\begin{equation}
\label{eq:meanresptimesinglestageRR}
\mathrm{E}[ R_{\text{RR}} ] = \frac{ n \rho_0 }{ \lambda } \left[ 1 + \, \frac{ \rho_0 }{ 1 - \rho_0 } \frac{ ( 1 + C_Y^2 ) ( C_A^2 / n + \rho_0^2 C_Y^2 ) }{ 2 ( C_A^2 / n + \rho_0^2 C_Y^2 ) } \right]
\end{equation}

\subsection{Least Work Left (LWL) dispatching}
\label{subsec:singlestageLWL}

In this case, the arrival stream of jobs is distributed across $n$ servers according to the anticipative policy \ac{LWL}.
Servers keep track of the amount of residual work required to clear their backlog at any given time.
Upon arrival of a new job, the dispatcher polls servers and collects the amount of unfinished work of each of them.
It then selects the server that has the least amount of unfinished work.
Ties are broken at random (e.g. when multiple servers are idle).

It can be shown that this arrangement is equivalent to a multi-server queuing system with $n$ servers and a single, centralized waiting line, i.e., a G/G/$n$ queue.
We resort then to Marchal's approximation for multi-server G/G queues.
The mean service time is $\mathrm{E}[ X ] = \mathrm{E}[ Y ] / \mu = n \, \rho_0 / \lambda$, 
the mean offered traffic is $A = \lambda \mathrm{E}[X] = \lambda \mathrm{E}[Y]/\mu = n \rho_0$ and the utilization coefficient of any server is $\rho =A / n = \rho_0$.
The mean response time is given by
\begin{align}
\label{eq:meanresptimesinglestageLWL}
\mathrm{E}[ R_{\text{LWL}} ] &= \mathrm{E}[ X ] + \mathrm{E}[ X ] \frac{ C( n, n \rho ) }{ n (1 - \rho ) }\phi  \nonumber \\
  &= \frac{ n \, \rho_0 }{ \lambda } \left[ 1 + \frac{ C( n, n \rho_0 ) }{ n (1 - \rho_0 ) } \frac{ ( 1 + C_Y^2 ) ( C_A^2 + \rho_0^2 C_Y^2 ) }{ 2 ( C_A^2 + \rho_0^2 C_Y^2 ) } \right]
\end{align}
where $C(m,A)$, for any positive integer $m$ and non-negative real $A$, is the Erlang-C formula given by
\begin{equation}
\label{eq:ErlangC}
C(m,A) = \frac{ B(m,A) }{ 1 - \frac{ A }{ m } + \frac{ A }{ m } B(m,A) }
\end{equation}
and $B(m,A)$ is the Erlang-B formula:
\begin{equation}
\label{eq:ErlangB}
B(m,A) = \frac{ \frac{ A^m }{ m! } }{ \sum_{ j = 0 }^{ m }{ \frac{ A^j }{ j! } } } = \begin{cases}
     1 & m = 0, \\
     \frac{ A B(m-1,A) }{ m + A B(m-1,A) } & m > 0.
\end{cases}
\end{equation}

\subsection{Join Idle Queue (JIQ) dispatching}
\label{subsec:singlestageJIQ}

The stateful, but not anticipative policy \ac{JIQ} works as follows.
Upon a new arrival, the dispatcher looks up a table of $n$ bits.
The $j$-th bit equal to 1 means that server $j$ is idle.
If there is at least one bit equal to 1 in the dispatcher table, the dispatcher picks one of them chosen uniformly at random and assigns the newly arrived job to the corresponding server.
The bit is reset.
If all bits are 0, the dispatcher selects one server at random among all $n$ servers.

On the server side, as soon as a server completes its last job and goes back to idle state it sends a message to the dispatcher, to set its bit to 1.

This policy requires a number of messages in the order of the number of jobs dealt with.
The complexity of dispatching is quite limited.
The price to pay is that \ac{JIQ} is shown to be sub-optimal, i.e., it is not Heavy-Traffic Delay Optimal \cite{Zhou2017}.
The reason is that, under heavy traffic, \ac{JIQ} essentially boils down to a flat random assignment of jobs to the cluster of $n$ servers, which is known (and intuitive) to be a sub-optimal policy.

Given this description of \ac{JIQ}, it appears that it behaves exactly as \ac{LWL} as long as there is at least an empty server.
If all servers are busy, \ac{JIQ} is equivalent to a random selection policy, which has a similar performance as \ac{RR}.
Given the number of servers, $n$, and the mean offered traffic $A = \lambda \mathrm{E}[X] = \lambda \mathrm{E}[Y] / \mu = n \rho_0$, the probability that all servers are busy can be approximated with the Erlang B formula $B(n, A) = B(n, n\rho_0)$.
Then, the mean response time can be approximated as follows:
\begin{equation}
\label{eq:meanresptimesinglestageJIQ}
\mathrm{E}[ R_{\text{JIQ}} ] \approx B( n , n\rho_0 ) \mathrm{E}[ R_{\text{RR}} ] + \left[ 1 - B( n , n \rho_0 ) \right] \mathrm{E}[ R_{\text{LWL}} ]
\end{equation}
where $B(m, A)$ is given in \cref{eq:ErlangB} and subscripts on the variable $R$ highlight the policy to be considered for computing the relevant mean response time.

%

\section{Performance evaluation}
\label{sec:numres}

As mentioned in \cref{sec:dataset}, each job consists of one or more tasks. 
In the analysis of the dispatching algorithms, we carry out the performance evaluation in two scenarios:
\begin{itemize}
    \item Job level -- Ignore the task level of the data and consider only jobs, using the aggregated service time of all tasks belonging to each job.
    \item Consider the task level of the data, allowing independent assignment of tasks to the server.
\end{itemize}

Based on these two scenarios, we evaluated how dispatching policies, system architecture, and parallelism influence key performance metrics, such as mean job response time and job slowdown. 
The latter is defined as the ratio of the response time of the job to the job service time (the sum of service time of all tasks belonging to that job).
When analyzing performance at the job level, the slowdown is necessarily no less than 1.
In the task level analysis, thanks to the parallelism of multiple servers and independent task dispatching, it might be the case that the slowdown is even less than 1.

Simulations are run by extracting one day-worth workload data from the whole trace.
The first two moments of inter-arrival times (\acp{IAT}) and of service times (\ac{CPU} times) are estimated from the considered sample.


\cref{tab:parameters} provides an overview of the key parameters used in our data-driven simulations. 
To explore the impact of parallelism and architectural design on system performance, we maintained a fixed overall system capacity while varying the number $n$ of servers and the processing rate per server $\mu$. 
This was done under the constraint that the product $n \mu$ remained constant, ensuring adherence to the fixed utilization coefficient $\rho_0$. 

\begin{table}[]
\centering
\caption{Simulation parameters}
\label{tab:parameters}
\resizebox{\columnwidth}{!}{%
\begin{tabular}{|l|l|}
\hline
\textbf{Parameter}             & \textbf{Value/Description}                   \\ \hline
Number of Servers, $n$    & [2, 1000]  \\ \hline
Target utilization coefficient of servers, $\rho_0$   &  0.8	 	\\ \hline
Server Processing Rate, $\mu$  &  Computed based on fixed overall capacity     \\ \hline
Threshold, $\theta$           &  \{10, 20, 30, 40, 50, 60, 70, 80, 90, 99\}\\
                                          &  quantile of task CPU time    \\ \hline
Dispatching Policies         & RR, JIQ, LWL                           \\ \hline
Scheduling Policy             & FCFS                                         \\ \hline
Performance Metrics        & Mean Response Time, Mean Slowdown \\ \hline
\end{tabular}%
}
\vspace{-0.35cm}

\end{table}

These parameters were chosen to reflect realistic scenarios commonly found in large-scale computing environments. 
By varying the number of servers, we assessed the scalability of dispatching policies under increasing parallelism, while adjustments to the processing rate per server provided insights into the trade-offs between server capacity and load distribution.
Our evaluation focused on three dispatching algorithms: \ac{RR}, \ac{JIQ}, and \ac{LWL}, combined with \ac{FCFS} scheduling across all servers. 

\subsection{Job-level performance evaluation in a single-stage cluster}

In this subsection, we focused exclusively on job-level data to evaluate the performance of each dispatching policy using both analytical models and data-driven simulations. 
%

\begin{figure}[t]
	\centering
	\includegraphics[width=0.8\columnwidth, trim=0cm 0cm 0cm 0cm, clip]{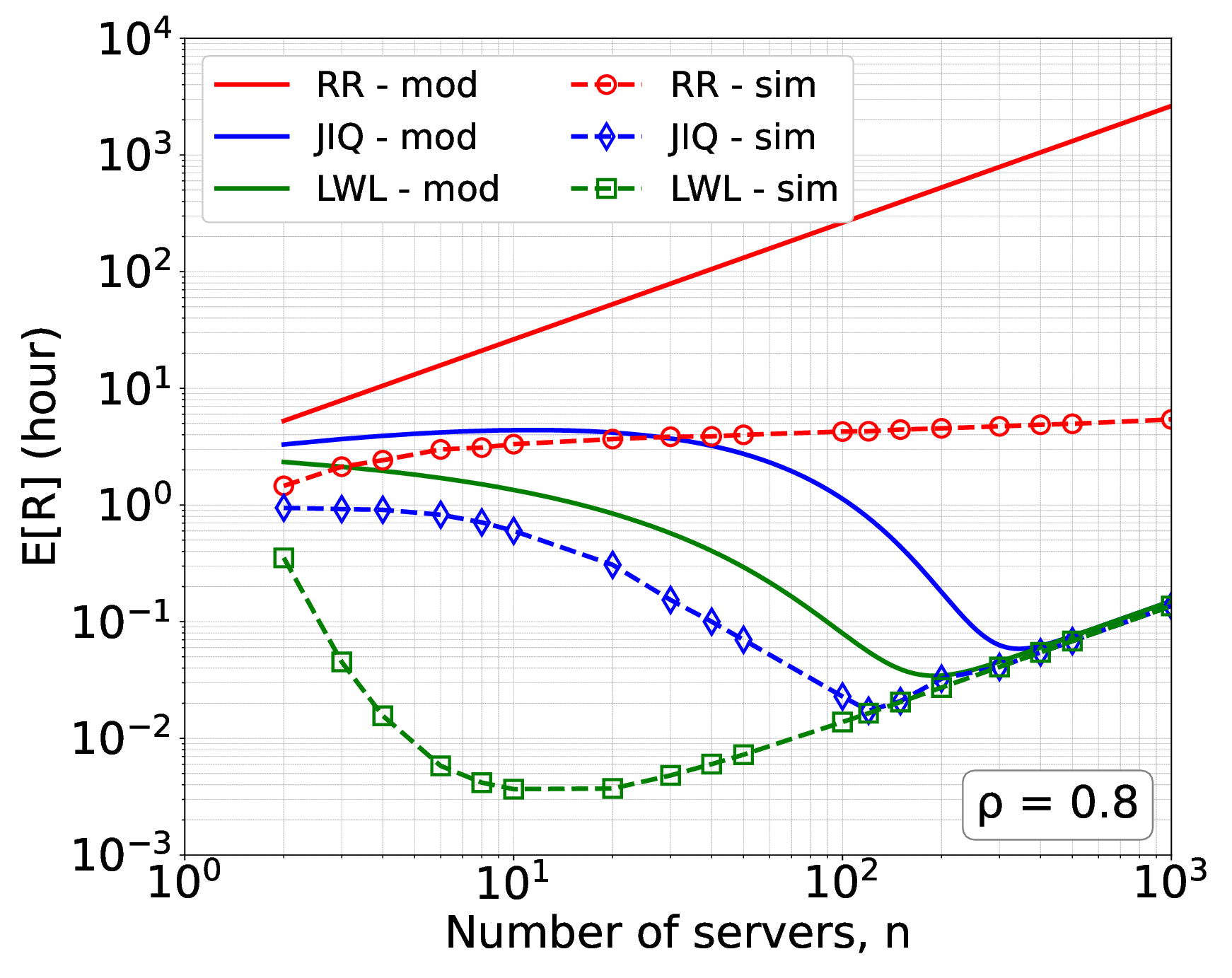}
	\caption{Mean response time of a cluster of $n$ servers as a function of $n$. Lines are computed by using analytical models, and markers with the dashed lines correspond to data-driven simulations based on the workload of day 4 of the Google trace of cell c. The utilization coefficient of servers is $\rho_0 = 0.8$. \textbf{Note: The data used includes original values with outliers and is not shuffled.}}
	\label{fig:modsimnorm}

\vspace{-0.35cm}
\end{figure}

\Cref{fig:modsimnorm} shows job mean response time as a function of the number $n$ of servers for the original job-level data and a target utilization coefficient $\rho_0 = 0.8$.
By ``original'' we mean that the jobs, their arrival times, and CPU demands were considered exactly as provided in the dataset.  
Solid lines in \Cref{fig:modsimnorm}  are obtained by means for the analytical models introduced in \cref{sec:modeldescription}.
Simulation results are illustrated as markers, connected with a dashed line.
Marker style and colors are preserved throughout the plots of this Section.
 
It is apparent that models fail to give accurate predictions of performance.
Yet, the qualitative behavior of the dispatching policies is correctly predicted by models.
Namely, \ac{RR} yields a monotonously increasing job response time as the number of servers grows.
Due to poor job dispatching, \ac{RR} does not benefit from increased parallelism and only suffers from decreasing server speed as $n$ grows.
On the contrary, both \ac{JIQ} and \ac{LWL} exhibit a minimum for some optimal degree of parallelism.
This optimal configuration stems from striking the best trade-off between a high degree of parallelism, which helps prevent a large job to stuck in the queue and imposes a high delay on several smaller jobs, and the speed of each server, which decreases as $n$ grows.
For small values of $n$, a few big jobs are already enough essentially to choke the server cluster.
For very large values of $n$, almost no jobs have to wait before a server is assigned, but servers are very slow.
Hence, there is an optimal degree of parallelism.

\ac{LWL} turns out to offer the best performance among the three considered policies, which is not surprising in view of the fact that this is also the most demanding policy.
It requires knowledge of job size as well as of the state of the servers.
As expected, the sub-optimal policy \ac{JIQ} yields worse performance as compared to \ac{LWL}.
Yet, for sufficiently large values of $n$, \ac{LWL} and \ac{JIQ} boils down to essentially the same policy, since there is an idle server upon each job arrival with high probability (see \cref{fig:Probidle}).
This remark points at a very interesting outcome of this analysis.
A sophisticated policy as \ac{LWL} brings no advantage over a much simpler one, as \ac{JIQ}, if the system architecture is properly designed.
In this case, we simply need to boost the parallelism of the server cluster large enough.
However, deploying a large number of slow servers, while making \ac{LWL} and \ac{JIQ} equivalent, does not yield the best possible performance at the job level.
We will dig more along these lines in the next section, with the task level analysis. 

Summing up, \ac{RR} mean job response time grows steadily with $n$, while \ac{JIQ} and \ac{LWL} have a minimum for some optimal $n$, \ac{LWL} offering uniformly better performance as compared to \ac{JIQ}.
The same remarks can be made based on the results of analytical models.
Under this respect, we may say that models provide a correct \emph{qualitative} insight, even if numerical predictions are grossly departing from simulations.

To investigate the reasons behind the observed mismatch, we analyzed the impact of \ac{IAT}, \ac{CPU} time, and the influence of outlier jobs. 
For this purpose, we consider the following modifications of workload data used to drive simulations.
\begin{itemize}
    \item Shuffle the \acp{IAT} and leave the sequence of job \ac{CPU} demands unchanged.
    \item Shuffle job \ac{CPU} demands and leave the sequence of \acp{IAT} unchanged.
    \item Maintain the sequences of \acp{IAT} and job \ac{CPU} demand unchanged, but remove those job whose \ac{CPU} demand is larger than the 99.9\% quantile of \ac{CPU} demand (\emph{outlier jobs}).
\end{itemize}

The first two modifications are motivated by the aim to break any possible correlation structure in the measured workload traces.

\begin{figure*}[t]
    \centering
    \subfigure[Original data.]{\includegraphics[width=.31\textwidth]{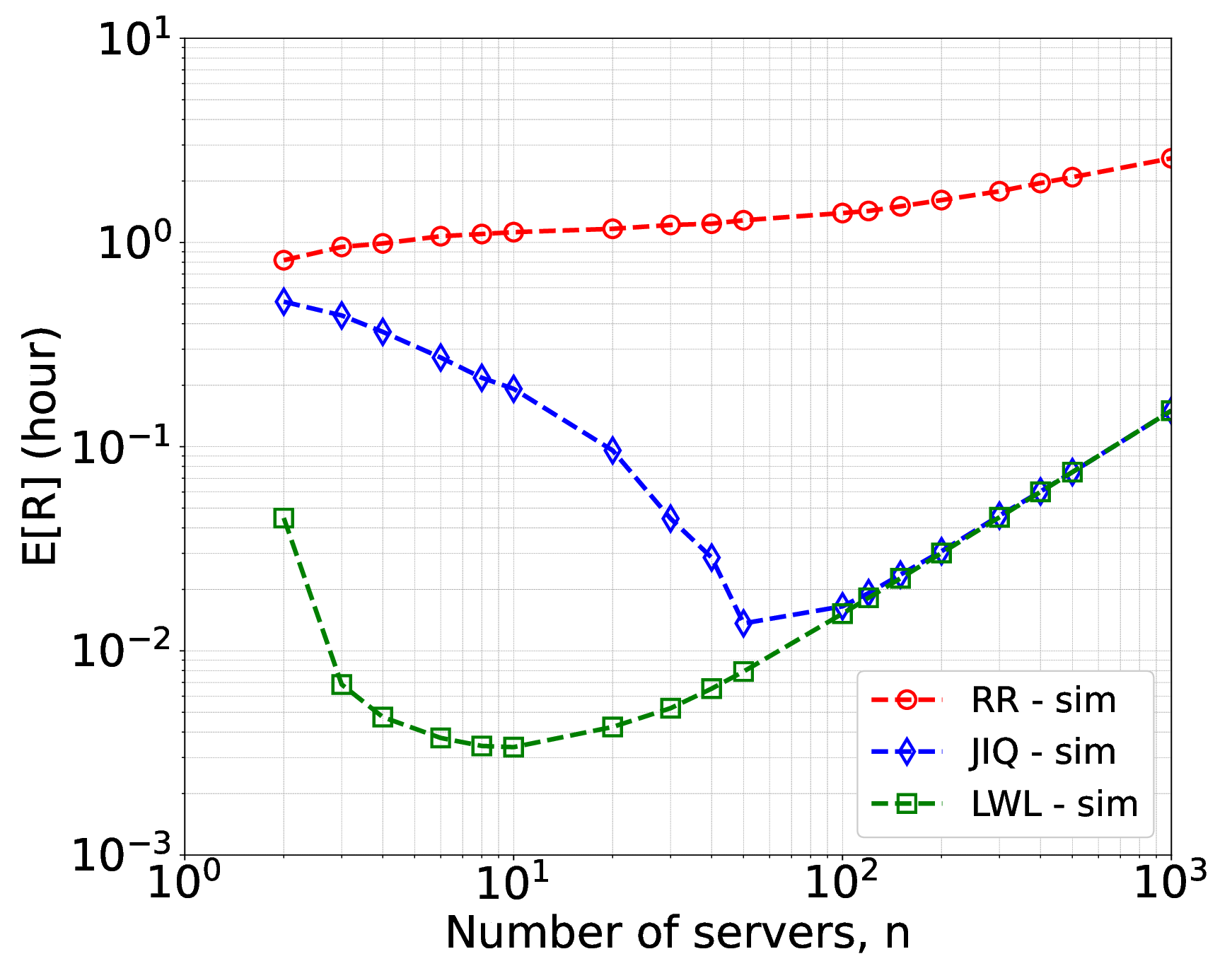}\label{fig:Jobsnormal}}\;
    \subfigure[Only \acp{IAT} shuffled.]{\includegraphics[width=.31\textwidth]{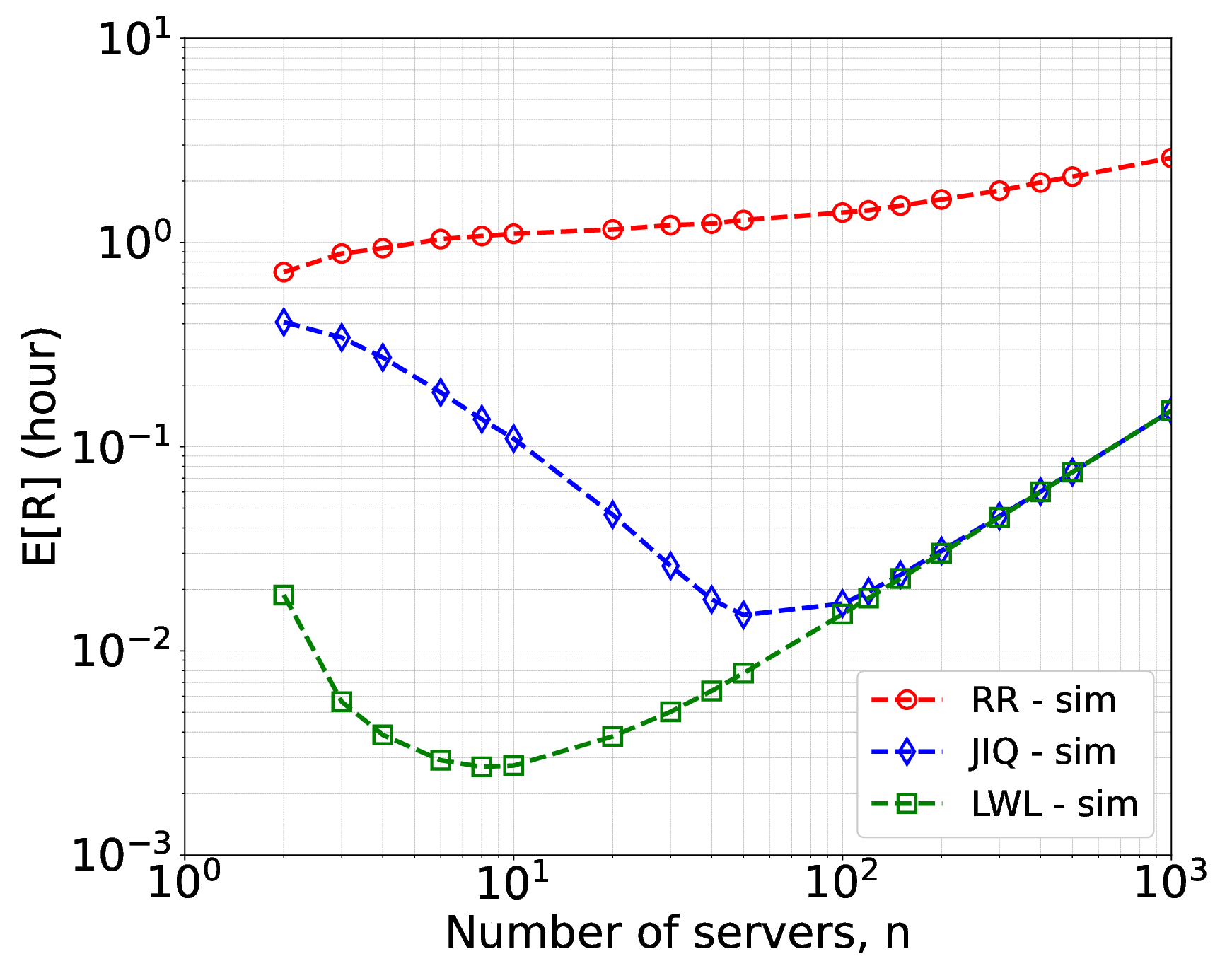}\label{fig:JobsshuffleIAT}}\;
    \subfigure[Only \ac{CPU} shuffled.]{\includegraphics[width=.31\textwidth]{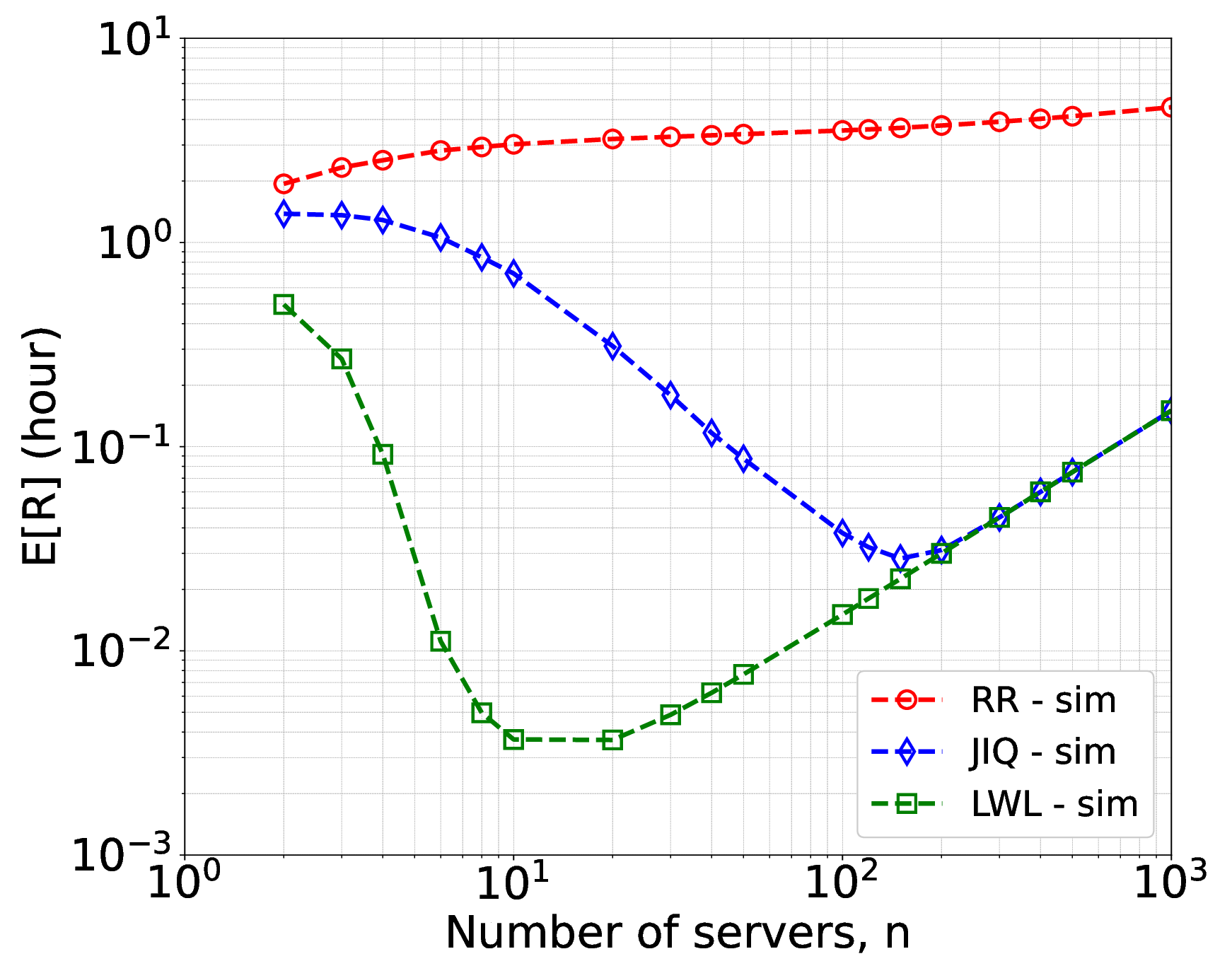}\label{fig:JobsshuffleCPU}}\;
    \caption{Decomposition of original data and comparative analysis of mean response time of a cluster of $n$ servers as a function of $n$ under different conditions.}
    \label{fig:jobs_analysis}

\vspace{-0.35cm}
\end{figure*}

\cref{fig:jobs_analysis} compares the mean job response time as a function of the number of servers $n$ for the original data (\cref{fig:Jobsnormal}) and for the first two modifications (\cref{fig:JobsshuffleIAT,fig:JobsshuffleCPU}, respectively).

From these experiments, we observed that \ac{IAT} has a negligible effect on performance. 
However, shuffling the \ac{CPU} demands significantly influences the performance of all dispatching policies. 

\begin{figure}[t]
    \centering
    \includegraphics[width=.8\columnwidth]{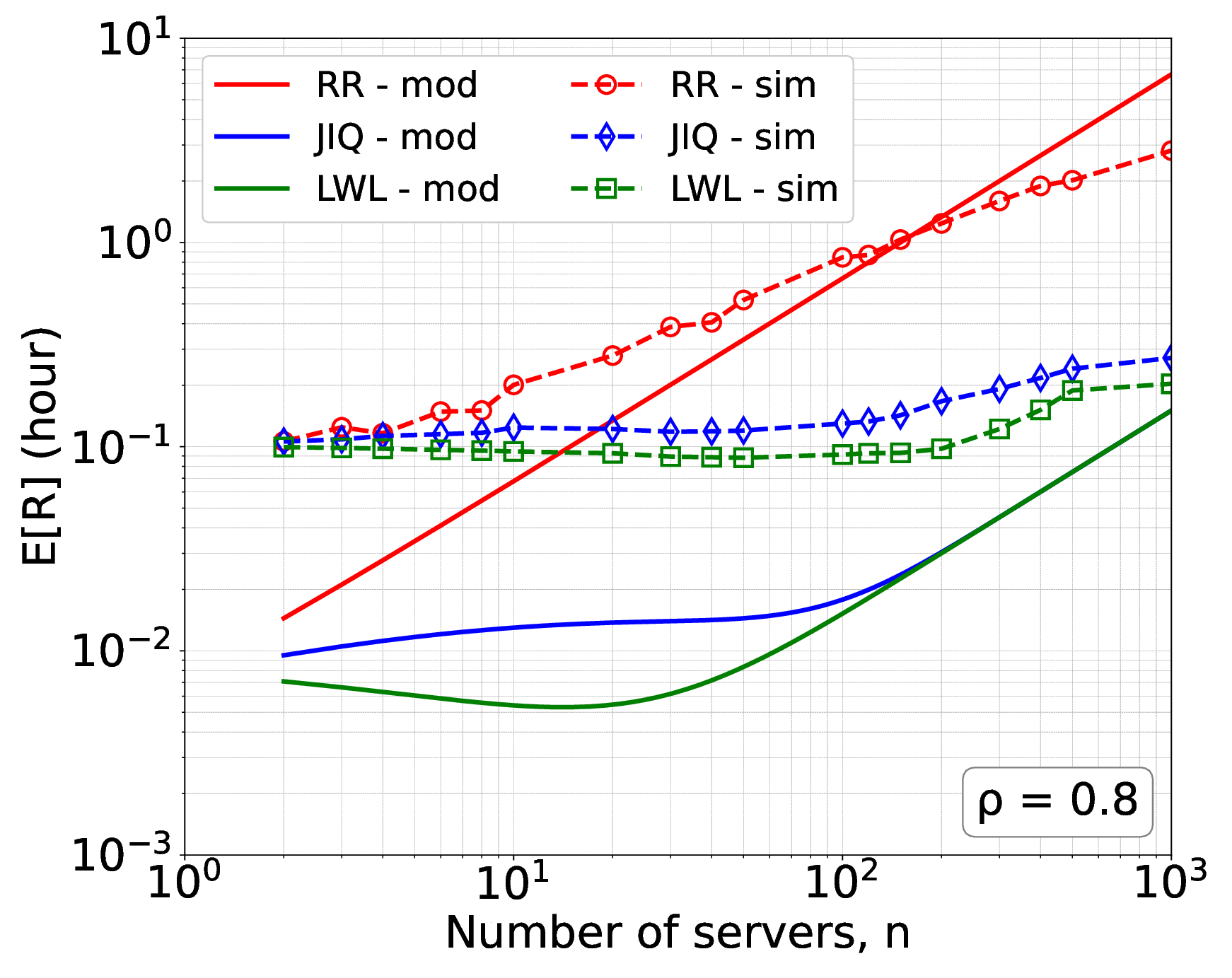}
    \caption{Mean response time of a cluster of $n$ servers as a function of $n$. Lines are computed by using analytical models, and markers with the dashed lines correspond to data-driven simulations based on the workload of day 4 of the Google trace of cell c. The utilization coefficient of servers is $\rho_0 = 0.8$. \textbf{Note: The data used does not include outliers but both \ac{IAT} and \ac{CPU} values are kept unchanged.}}
    \label{fig:modelsimnoout}
\vspace{-0.35cm}
\end{figure}

\cref{fig:modelsimnoout} plots the mean job response time as a function of $n$ in case of the third workload modification listed above, i.e., outlier jobs have been removed.

Compared with previous results, it is seen that removing just 0.01\% of jobs with the largest \ac{CPU} time values (outliers) has a substantial impact on the mean response time, highlighting the critical role of these "monster" jobs in overall system performance.
While maintaining the same qualitative behavior, analytical models still fail to capture the performance of data-driven simulations.

\begin{figure}[t]
	\centering
	\includegraphics[width=.8\columnwidth]{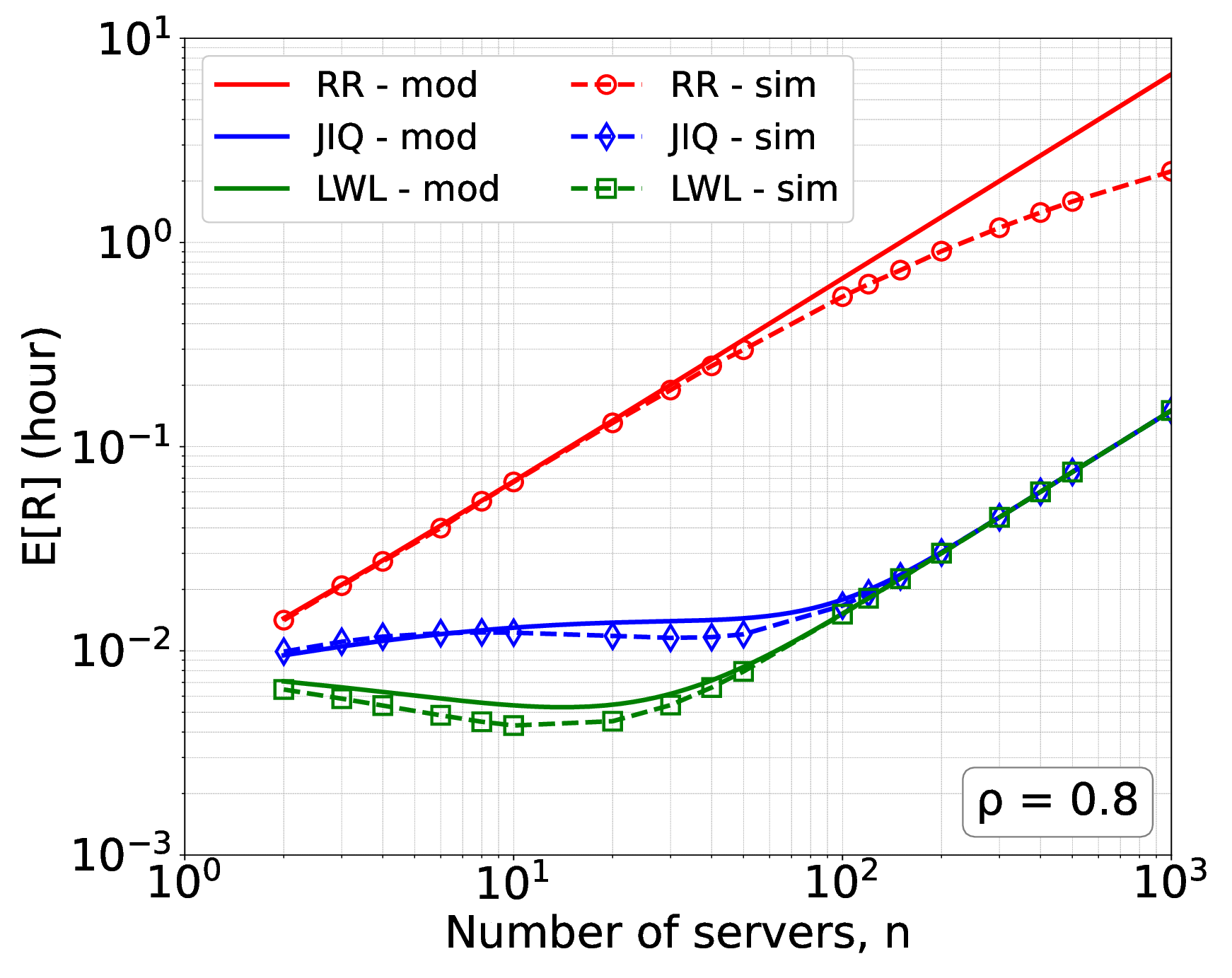}
	\caption{Mean response time of a cluster of $n$ servers as a function of $n$. Lines are computed by using analytical models, and markers with the dashed lines correspond to data-driven simulations based on the workload of day 4 of the Google trace of cell c. The utilization coefficient of servers is $\rho_0 = 0.8$. \textbf{Note: The data used does not include outliers and both \ac{IAT} and \ac{CPU} values are shuffled.}}
	\label{fig:modelsimmatch}
\vspace{-0.35cm}
\end{figure}

Finally, we apply all modifications simultaneously, i.e., we construct a workload trace by removing outliers from the original data, then shuffling both \acp{IAT} and \ac{CPU} times, to break any correlation in the original data sequences.
The corresponding results are shown in \cref{fig:modelsimmatch}, where the job mean response time is plotted versus $n$.

This time analytical models appear to be in excellent agreement with data-driven simulation results.
Further analysis, not shown for the sake of space, shows that restoring the original sequence of \acp{IAT} does not affect the accuracy of the model significantly.
These results point out that the discrepancy between model predictions and simulation results are rooted in two key characteristics of the workload traces:
\begin{itemize}
	\item \ac{CPU} times bear a correlation structure that strongly impacts performance. The analytical models fail since they are based on renewal assumptions that disregard any correlations among service times.
	\item Outlier jobs introduce a distortion in matching the first two moments of service times that reflects in inaccurate predictions of the model.
\end{itemize}
	


\begin{figure}[ht]
\vspace{-0.35cm}
	\centering
	\includegraphics[width=.95\columnwidth]{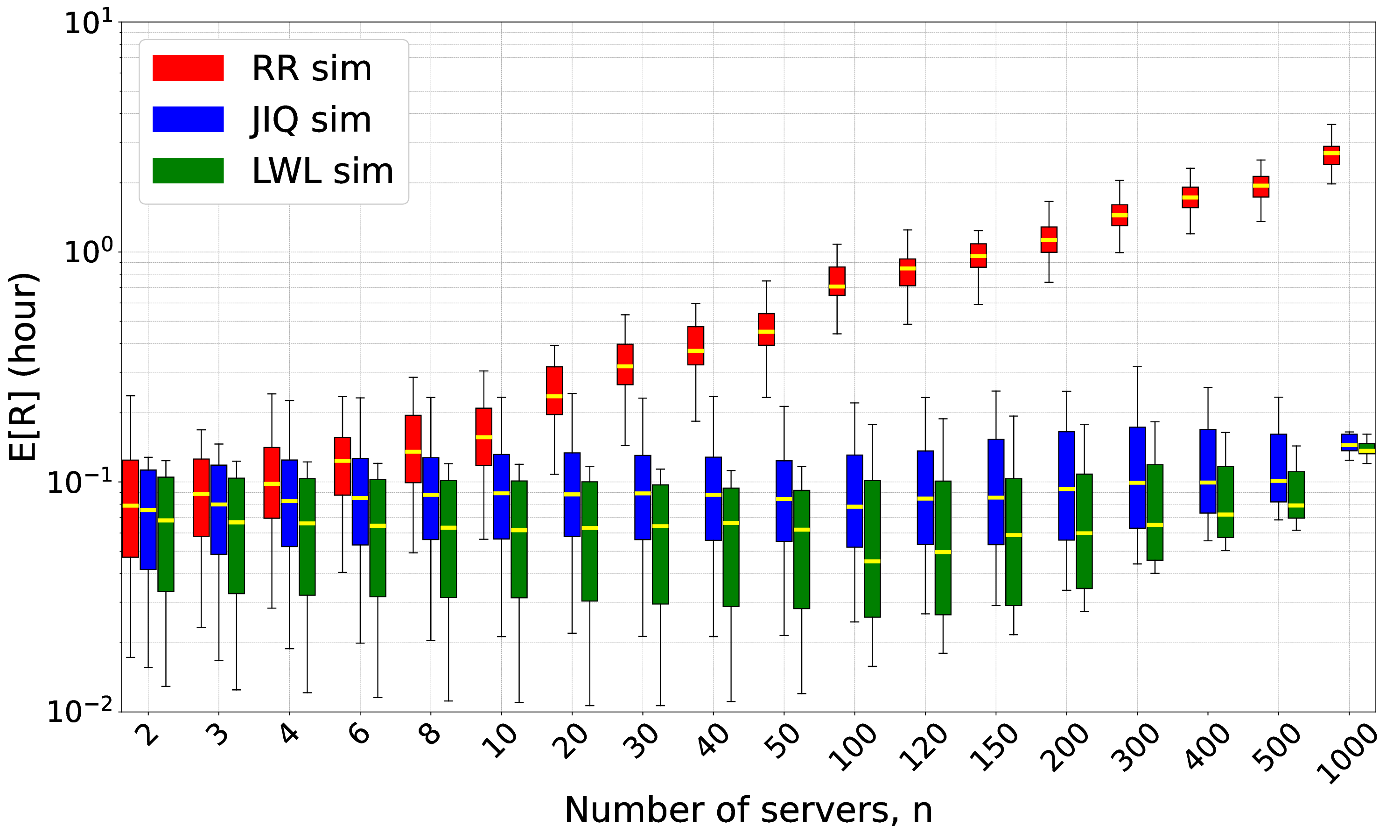}
	\caption{Mean response time of dispatching algorithms for varying server counts over 31 days in May, shown as boxplots with the yellow line indicating the median of the mean response time. The boxplots visualize the spread.}
	\label{fig:monthlyspread}
\vspace{-0.35cm}
\end{figure}

The whole analysis above is based on one-day-worth of the workload traces.
We therefore examined whether there is a significant difference in performance across different days of the month. 
For this purpose, we evaluate the mean job response time for each of the 31 days of the original trace and present the results in \Cref{fig:monthlyspread}, as a function of the number of servers $n$.
The diagram in \cref{fig:monthlyspread} is a box-plot, showing the median (yellow dash inside the vertical rectangles), the interval covered by those samples that lie between the 25\% and the 75\% quantiles of the obtained results day by day (vertical rectangle) and the entire range of the sample (error-bar).
It is apparent that there is some spread of obtained mean job response time day by day, but this does not affect any of the remarks we have made on the behavior of the mean response time with the three considered dispatching policies.
It is to be noted that \ac{RR} appears to be more stable than the other two policies.

The results presented in \cref{fig:monthlyspread} indicate that the analyses conducted so far are consistent across different days. 
In other words, the data exhibits similar behavior regardless of the day being analyzed.

\begin{figure}[t]
    \centering
   \includegraphics[width=.8\columnwidth]{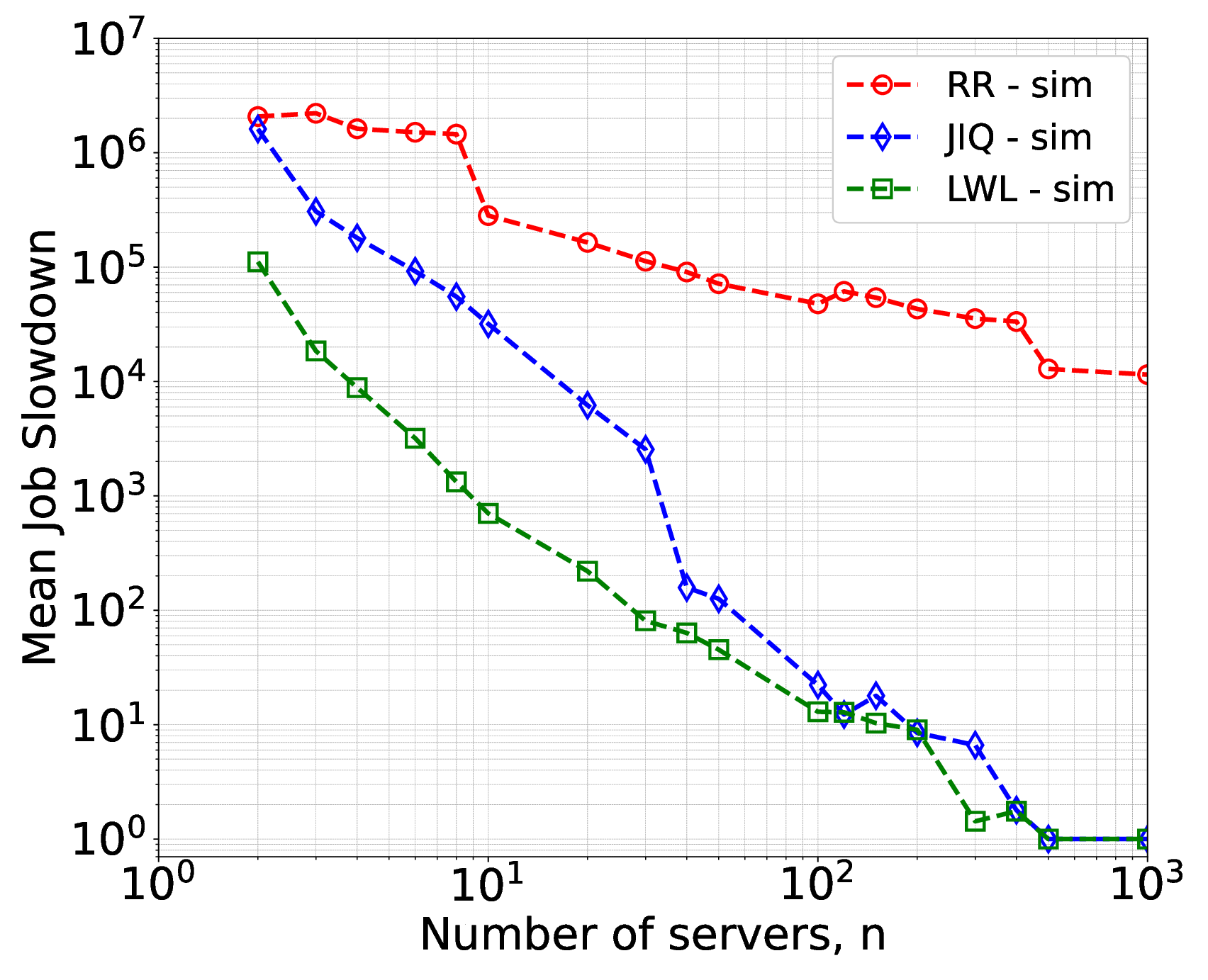}
    \caption{Job slowdown of a single-stage server cluster as a function of the number of servers $n$ for job level dispatching.}
    \label{fig:JobsSD}

\vspace{-0.35cm}
\end{figure}

We further analyze the slowdown behavior with job level dispatching in \cref{fig:JobsSD}. 
For job-level dispatching, as expected, the \ac{LWL} policy consistently outperformed \ac{JIQ}. 
Performance of \ac{LWL} and \ac{JIQ} converge under high parallelism. 
In such scenarios, both policies behave similarly because an idle server is almost always available when a job arrives, allowing both to assign the job to the idle server, resulting in a slowdown equal to 1. 

%

\subsection{Task-level performance evaluation in a single-stage cluster}

Moving to the more realistic scenario where the internal structure of jobs is exploited for a more flexible assignment of workload, we assume tasks belonging to the same job can be dispatched to servers independently of one another.
The considered metric is still mean job response time, where a job is completed only once all of its tasks are done.

\begin{figure*}[t]
    \centering
    \subfigure[Original data.]
     {\includegraphics[width=.31\textwidth]{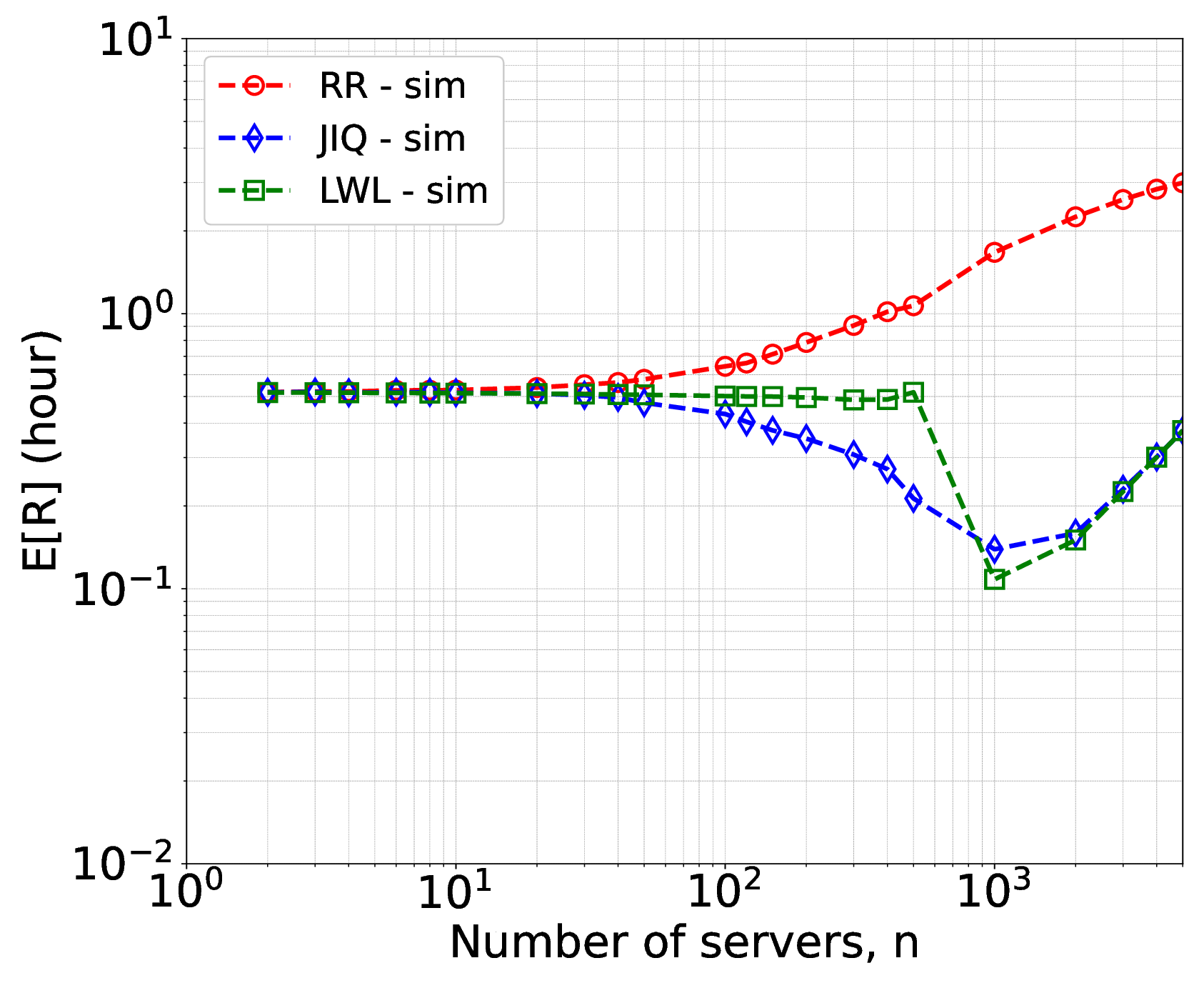} \label{fig:Tasknormal}}\;
     \subfigure[Only \acp{IAT} shuffled.]{\includegraphics[width=.31\textwidth]{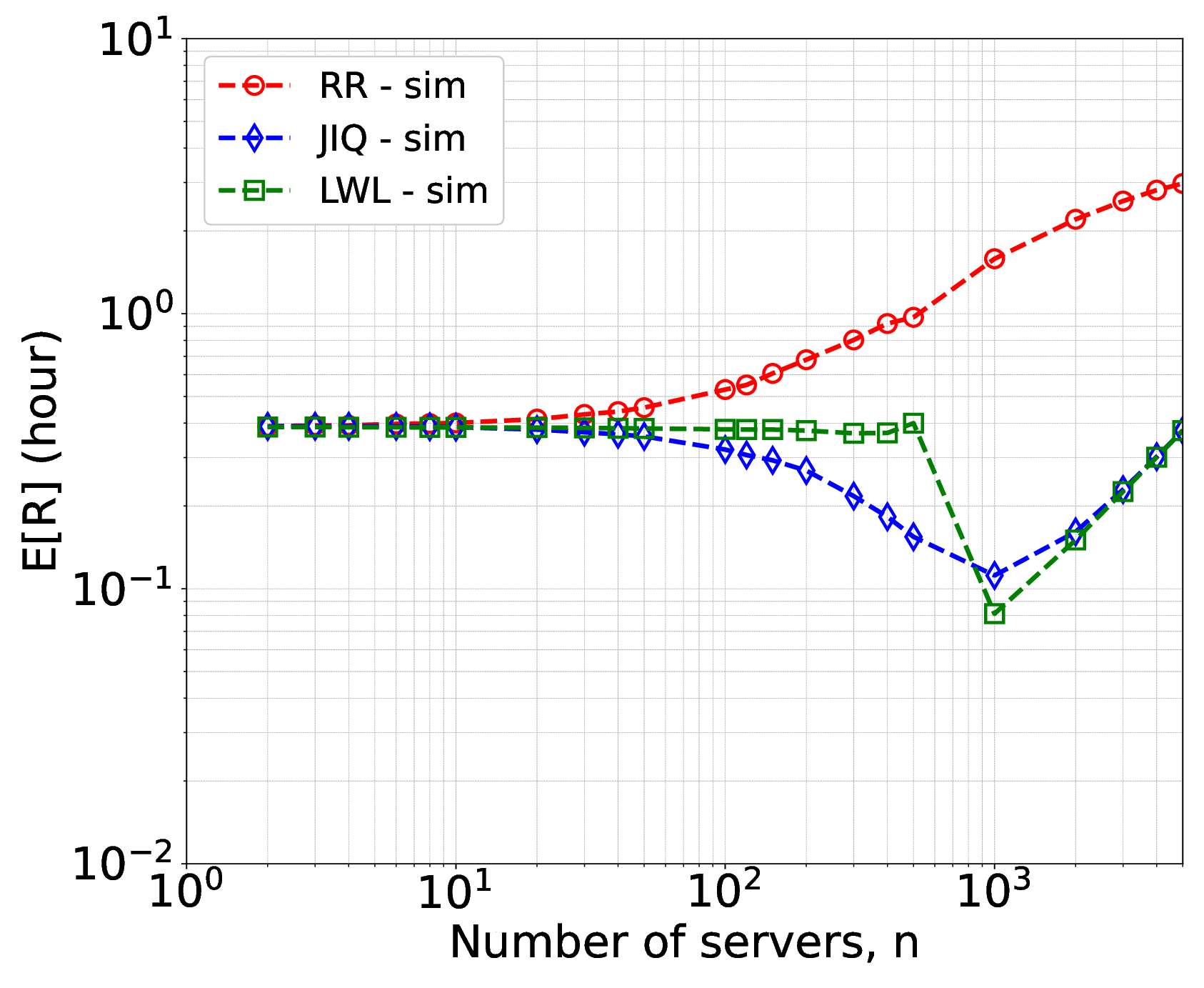}\label{fig:TaskshuffleIAT}}\;
     \subfigure[Only \ac{CPU} times shuffled.]{\includegraphics[width=.31\textwidth]{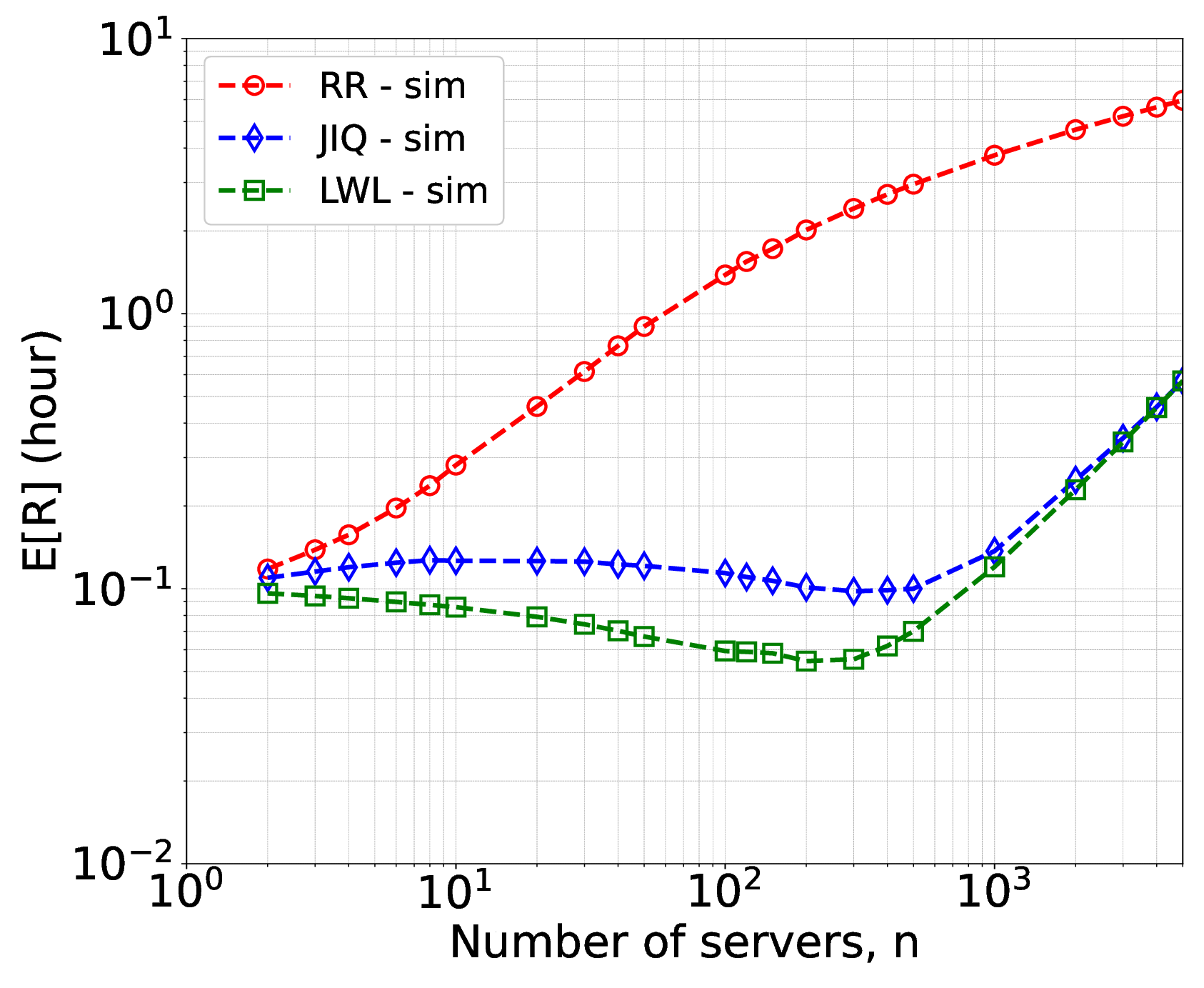}\label{fig:TasksuffleCPU}}\;
    \caption{Decomposition of original data and comparative analysis of mean response time of a cluster of $n$ servers as a function of $n$ under different conditions. The data structure includes task level but E[R] is the mean job response time.}
    \label{fig:tasks_analysis}

\vspace{-0.35cm}
\end{figure*}

\cref{fig:Tasknormal} shows the mean job response time as a function of $n$ for the task-level system.
Mean response times are quite different from the case of job-level analysis.

Most interestingly, for a wide range of the number of servers, \ac{JIQ} turns out to be the best policy, beating the supposedly more performing \ac{LWL}.
As already seen in the job-level analysis, for very large values of $n$ the two policies tend to coincide.
As for \ac{RR}, it is still monotonously increasing with $n$, showing that this simple stateless policy cannot benefit from parallelism in a single-stage server cluster.

The best-performing configuration is attained from $n$ in the order of several hundred servers.
At the optimum working point, \ac{LWL} and \ac{JIQ} offer the same performance level, once again proving that it is not necessary to resort to a size-based (anticipative) policy.

To dig further in the comparison among these policies with task level dispatching, the effect of shuffling either \acp{IAT} of \ac{CPU} times\footnote{\ac{CPU} times are shuffled among all tasks, yet preserving the number of tasks belonging to each job.}
 is shown in \cref{fig:TaskshuffleIAT,fig:TasksuffleCPU} respectively.
It can be observed that shuffling \acp{IAT} has a negligible effect on performance compared to those obtained with the original workload.
On the contrary, shuffling \ac{CPU} times significantly alters the simulation results.
\ac{LWL} is restored as the winning policy, thus aligning with results found with job-level dispatching.
This result points out the root cause for the surprising policy ranking in \cref{fig:Tasknormal}: correlation structure of \ac{CPU} times, giving rise to ``monster'' jobs that comprise a large number of big tasks.

More in depth, we believe the reason for the behavior shown in \cref{fig:tasks_analysis} lies in how \ac{LWL} balances the workload across all servers. 
There are jobs with thousands of tasks, and when these large jobs arrive, \ac{LWL} evenly distributes tasks to servers, thus ultimately blocking all servers. 
This causes smaller jobs, which have a single task (the vast majority of jobs is that way), to get stuck in the system while waiting for the large jobs to complete. 

On the other hand, \ac{JIQ} behaves differently when handling large jobs with thousands of tasks. 
The policy assigns tasks one by one to idle servers and, once no idle servers are available, starts assigning tasks randomly to busy servers. 
We believe this behavior benefits smaller jobs and tasks, as it prevents that \emph{all} servers from being completely blocked by the large jobs.
Unequal server load favors some servers becoming idle again sooner than it would happen with \ac{LWL}.

This behavior can be illustrated by the following argument.
Say the $n$ servers are all busy and let $U_1, U_2, \dots, U_n$ be the unfinished work in the $n$ servers.
For ease of notation, let us assume that the $U_k$
s are sorted in ascending order, i.e., $U_1 \le U_2 \le \dots \le U_n$.
Let $X$ be the service time of a task to be dispatched.
With \ac{LWL}, the task goes to server 1.
Hence, the minimum unfinished work among all servers, \emph{after} the new task has been dispatched, is $U_{\text{LWL}} = \min\{ U_1+X, U_2 \}$.
In the case of \ac{JIQ}, with probability $1/n$ the task is dispatched to server 1 and the outcome is the same as with \ac{LWL}.
With probability $1-1/n$, the task goes to any other server.
In the latter case, the minimum unfinished work, \emph{after} the new task has been dispatched, is $U_{\text{JIQ}} = U_1$.
Hence, $U_{\text{LWL}}  - U_{\text{JIQ}}  = 0$ with probability $1/n$, while $U_{\text{LWL}}  - U_{\text{JIQ}}  = \min\{ U_1+X, U_2 \} - U_1 = \min\{ X, U_2 - U_1 \} \ge 0$ with probability $1-1/n$.
For large $n$ it is seen that $U_{\text{LWL}} > U_{\text{JIQ}}$ with high probability.
This implies that servers are biased to becoming idle again sooner with \ac{JIQ} rather than with \ac{LWL}.
Getting back an idle server sooner benefits other jobs, especially smaller ones, made up of a single task.
 
\begin{figure}[t]
	\centering
	\includegraphics[width=.8\columnwidth]{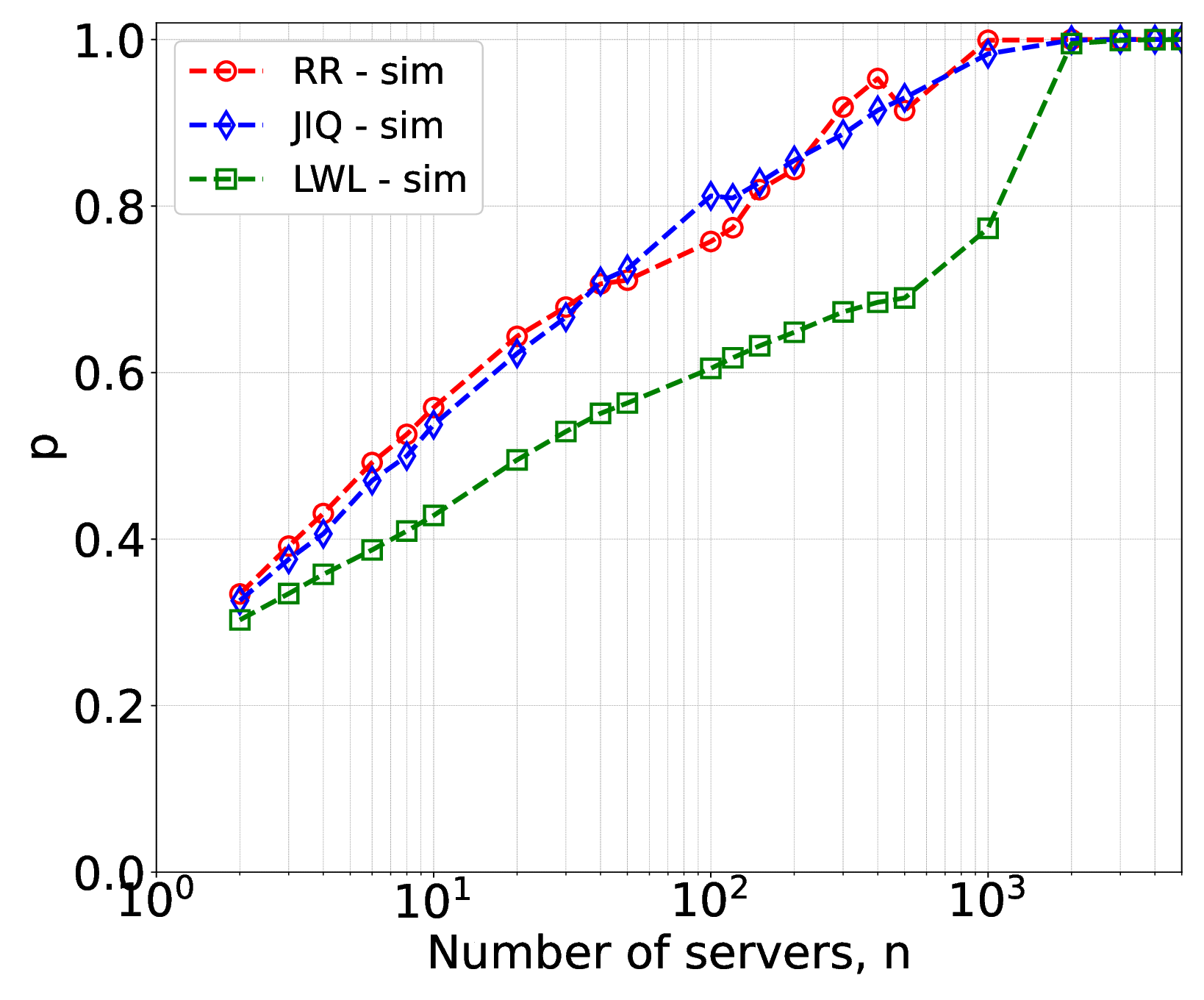}
	\caption{Probability of at least one server being idle upon job arrival, shown as a function of the number of servers for various dispatching policies.
    \textbf{Note: The task level data structure has been used.}}
	\label{fig:Probidle}
    
\vspace{-0.35cm}
\end{figure}

To further support our hypothesis, we analyze the probability of having at least one idle server upon job arrival, which is plotted in \cref{fig:Probidle} against $n$.
It is apparent that this event is quite more probable with \ac{JIQ} than with \ac{LWL}, which tends to equalize the backlog of all servers.

\begin{figure}[t]
    \centering
    \includegraphics[width=.8\columnwidth]{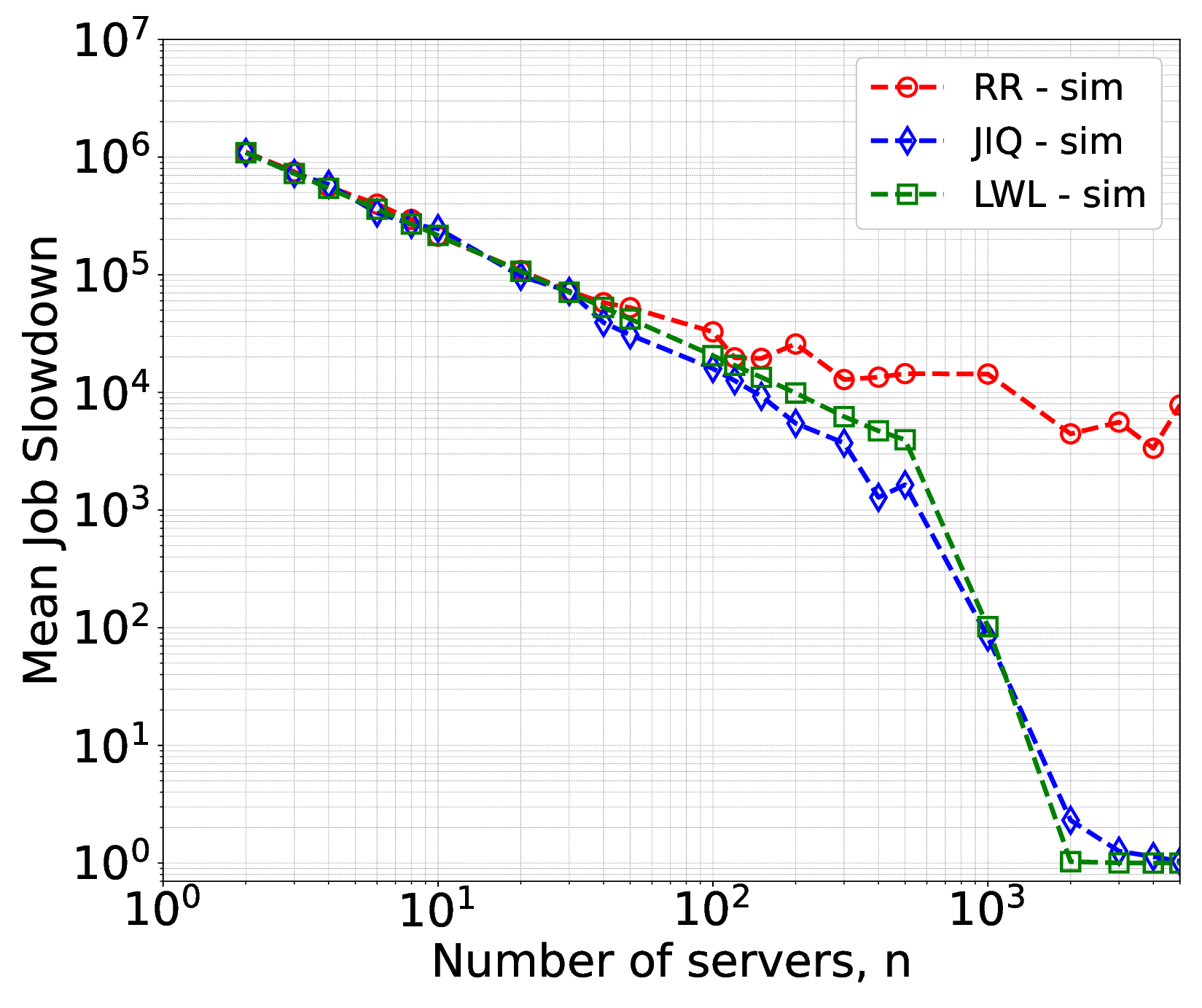}
    \caption{Job slowdown in a single-stage server cluster as a function of $n$ with task level dispatching.}
    \label{fig:TasksSD}

\vspace{-0.35cm}
\end{figure}

For very large values of $n$ (beyond about 500 servers), the probability of finding an idle server is essentially 1 for both \ac{JIQ} and \ac{LWL}.
This is the region where the two policies yield the same mean job response time.

Summing up, the randomness of task assignment with \ac{JIQ} makes server backlog unequal on a short-time scale, which favors some servers to become idle again sooner than with \ac{LWL}.
Note however that pure random assignment would yield unsatisfactory performance as proved by looking at \ac{RR}.
\ac{JIQ} introduces a minimum of smartness in choosing an idle server as long as there is one.
When all of them are busy if $n$ is large enough, randomly assigning tasks turns out to outperform more ``aware'' assignments.

Finally, the job slowdown with task level dispatching is plotted in \cref{fig:TasksSD} against $n$.
In contrast to the case of job-level dispatching, the task-level results reveal a different trend. 
While \ac{JIQ} generally achieves a lower mean response time than \ac{LWL}, it turns out that \ac{LWL} yields superior slowdown performance with respect to \ac{JIQ}.
This is not in contradiction with the results shown for the mean response time, in view of the intuitive explanation given above.
It is true that \ac{LWL} does a better job than \ac{JIQ} in distributing the workload evenly, and this brings better slowdown performance.
However, in the presence of ``monster'' jobs with a large number of heavy tasks, it is just this equalization of server backlog that makes the server unfinished work grow uniformly, thus failing to give an escape channel to later smaller jobs.
It is an example of the curse of being too smart.

\subsection{Two-stage server cluster}

After observing the surprising results with the task-level data structure, where a simple policy like \ac{JIQ} outperforms a more advanced policy like \ac{LWL}, we were motivated to explore the impact of system architecture on response time performance under real-life workload data.  
Specifically, we questioned whether a suitable system design could allow even \ac{RR} to outperform smarter policies.
To explore this possibility, we adopted the two-stage dispatching and scheduling system proposed in \cite{yildiz2025merit}. 

The two-stage system operates as follows. 
When a task arrives, it is handled by a front-end dispatcher, which assigns it to one of the servers in the first stage, according to \ac{RR} policy. 
A service time threshold, $\theta$, is defined for the $n_1$ servers in this stage. 
If a task completes its processing within the threshold $\theta$, it exits the system, having been fully served. 
Tasks that exceed this threshold are stopped and transferred to the second stage dispatcher, again using \ac{RR} policy.
These tasks restart to be processed to completion on one of the servers in the second cluster, which comprises $n_2 = n - n_1$ servers.
All servers of both clusters adopt \ac{FCFS} scheduling.

In contrast to the system proposed in \cite{yildiz2025merit}, we conducted extensive hyperparameter tuning to optimize the server allocation between stages ($n_1$ and $n_2$) and to determine the optimal threshold $\theta$.
With this design, we play with architecture and meta-parameters, while adopting the simplest and most easily implementable policies.

\begin{figure}[]
	\centering
	\includegraphics[width=.8\columnwidth]{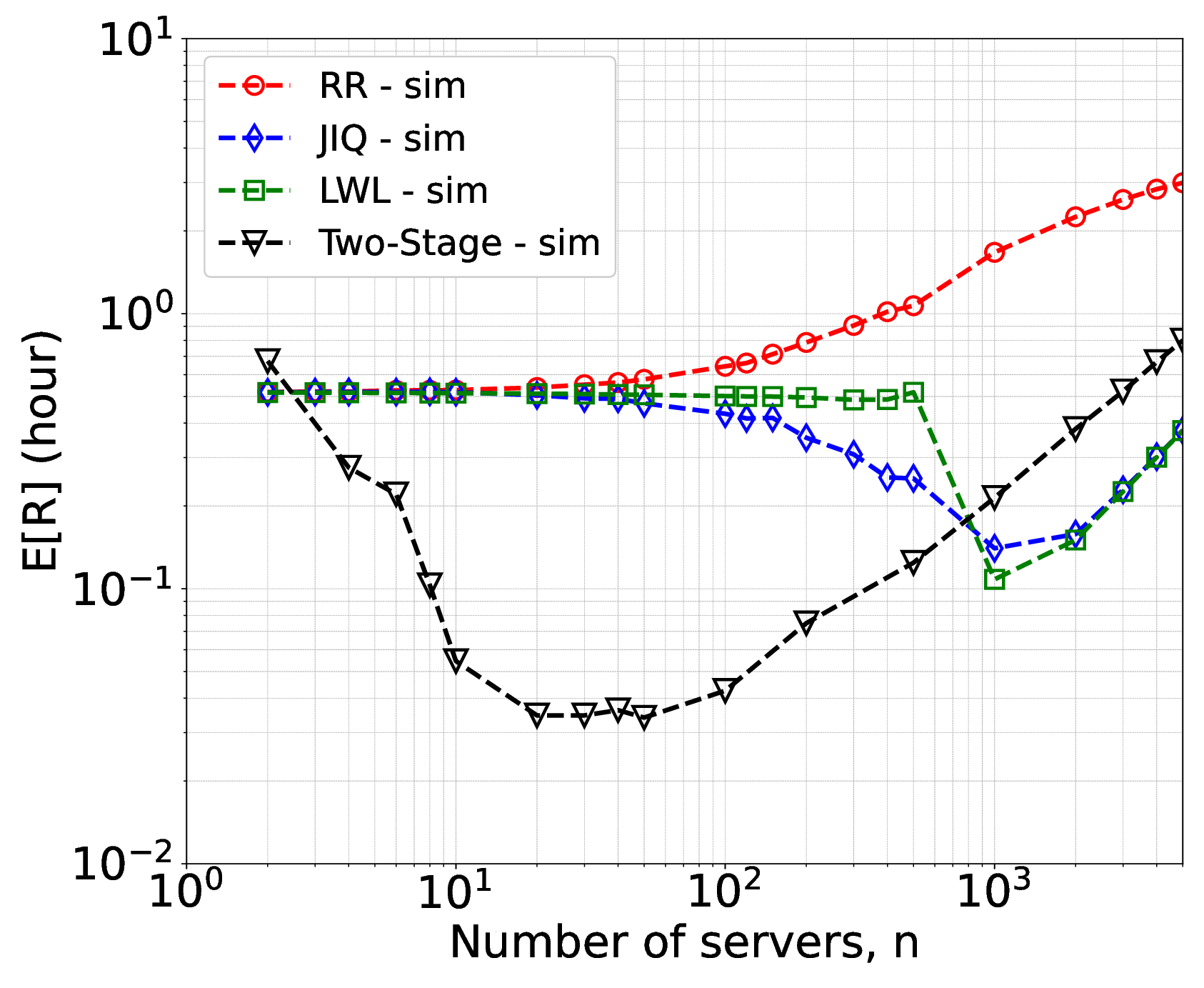}
	\caption{Mean job response time as a function of the overall number $n$ of servers for task level dispatching.}
	\label{fig:TwostageMRT}

\vspace{-0.35cm}
\end{figure}

Remarkably, despite its simplicity, the two-stage architecture outperforms single-stage multi-server systems, even adopting more sophisticated policies, such as \ac{LWL} and \ac{JIQ}.
This is illustrated in \cref{fig:TwostageMRT}, where the mean job response time is plotted against $n$ for task level dispatching, this time including also the two-stage architecture results.

The two-stage system achieves a lower mean response time with just 10 servers compared to a single-stage system with 1000 servers.  
By isolating short tasks ("mice") from longer ones ("elephants"), it effectively reduces delays and improves resource utilization.  
However, its performance declines with larger server counts, due to the decrease in server speed and the reduced number of servers in the first stage compared to the single-stage system, leading to diminished efficiency.  
Nevertheless, the two-stage system demonstrates the significant impact of architectural design on system performance, even when utilizing simple dispatching strategies.

%

\section{Conclusions}
\label{sec:conclusions}

In this paper, we analyzed several dispatching and scheduling policies in multi-server systems using real-world workload data. 
We highlighted key discrepancies between analytical models and simulations and demonstrated the limitations of traditional queuing theory in complex data center environments. 
Our results show that sophisticated policies like \ac{LWL} and \ac{JIQ} generally outperform simpler ones like \ac{RR}, though \ac{JIQ} can occasionally rival \ac{LWL} under specific workloads. 
Extending the analysis to task-level data revealed significant benefits in isolating short tasks ("mice") from longer ones ("elephants"), improving resource utilization and response times. 
The proposed two-stage system achieves remarkable performance improvements, outperforming single-stage configurations in scenarios with fewer servers. 
While its performance declines with larger server counts due to first-stage bottlenecks, the system highlights the impact of architectural design. 
Future work will focus on enhancing the scalability of the two-stage system and expanding its applicability to diverse workload types.

%
%

\section*{Acknowledgment}
This work was partially supported by the European Union under the Italian National Recovery and Resilience Plan (NRRP) of Next Generation EU, partnership on ``Telecommunications of the Future'' (PE00000001 - program ``RESTART'').

\bibliographystyle{IEEEtran}
\bibliography{references}

\begin{thebibliography}{10}
\providecommand{\url}[1]{#1}
\csname url@samestyle\endcsname
\providecommand{\newblock}{\relax}
\providecommand{\bibinfo}[2]{#2}
\providecommand{\BIBentrySTDinterwordspacing}{\spaceskip=0pt\relax}
\providecommand{\BIBentryALTinterwordstretchfactor}{4}
\providecommand{\BIBentryALTinterwordspacing}{\spaceskip=\fontdimen2\font plus
\BIBentryALTinterwordstretchfactor\fontdimen3\font minus \fontdimen4\font\relax}
\providecommand{\BIBforeignlanguage}[2]{{%
\expandafter\ifx\csname l@#1\endcsname\relax
\typeout{** WARNING: IEEEtran.bst: No hyphenation pattern has been}%
\typeout{** loaded for the language `#1'. Using the pattern for}%
\typeout{** the default language instead.}%
\else
\language=\csname l@#1\endcsname
\fi
#2}}
\providecommand{\BIBdecl}{\relax}
\BIBdecl

\bibitem{HarcholBalter2013}
M.~Harchol-Balter, \emph{Performance Modeling and Design of Computer Systems: Queueing Theory in Action}.\hskip 1em plus 0.5em minus 0.4em\relax Cambridge University Press, 2013.

\bibitem{Zhou2017}
X.~Zhou, F.~Wu, J.~Tan, Y.~Sun, and N.~Shroff, ``Designing low-complexity heavy-traffic delay-optimal load balancing schemes: Theory to algorithms,'' \emph{Proc. ACM Meas. Anal. Comput. Syst.}, vol.~1, no.~2, Dec. 2017.

\bibitem{KUMAR2019}
M.~Kumar, S.~Sharma, A.~Goel, and S.~Singh, ``A comprehensive survey for scheduling techniques in cloud computing,'' \emph{Journal of Network and Computer Applications}, vol. 143, pp. 1--33, 2019.

\bibitem{Tang2024}
S.~Tang, Y.~Yu, H.~Wang, G.~Wang, W.~Chen, Z.~Xu, S.~Guo, and W.~Gao, ``A survey on scheduling techniques in computing and network convergence,'' \emph{IEEE Communications Surveys \& Tutorials}, vol.~26, no.~1, pp. 160--195, 2024.

\bibitem{HarcholBalter2021}
M.~Harchol-Balter, ``Open problems in queueing theory inspired by datacenter computing,'' \emph{Queueing Syst. Theory Appl.}, vol.~97, no.~12, pp. 3--37, feb 2021.

\bibitem{Hyytia2020}
E.~Hyytiä and R.~Righter, ``Star and rats: Multi-level dispatching policies,'' in \emph{2020 32nd International Teletraffic Congress (ITC 32)}, 2020, pp. 81--89.

\bibitem{Choudhury2021}
T.~Choudhury, G.~Joshi, W.~Wang, and S.~Shakkottai, ``Job dispatching policies for queueing systems with unknown service rates,'' in \emph{Proceedings of the Twenty-second International Symposium on Theory, Algorithmic Foundations, and Protocol Design for Mobile Networks and Mobile Computing (MobiHoc ’21)}, 2021, pp. 181--190.

\bibitem{Bilenne2021}
O.~Bilenne, ``Dispatching to parallel servers,'' \emph{Queueing Systems}, vol.~99, no.~3, pp. 199--230, 2021.

\bibitem{AbdulJaleel2022}
J.~A. Jaleel, S.~Doroudi, K.~Gardner, and A.~Wickeham, ``A general 'power-of-d' dispatching framework for heterogeneous systems,'' \emph{Queueing Systems}, vol. 102, no. 3-4, pp. 431--480, 2022.

\bibitem{wilkes2019clusterdata}
J.~Wilkes, ``{cluster-data/Clusterdata2019},'' \url{https://github.com/google/cluster-data/tree/master}, 2019.

\bibitem{Tirmazi2020}
M.~Tirmazi, A.~Barker, N.~Deng, M.~E. Haque, Z.~G. Qin, S.~Hand, M.~Harchol-Balter, and J.~Wilkes, ``Borg: the next generation,'' in \emph{Proceedings of the Fifteenth European Conference on Computer Systems}, ser. EuroSys '20.\hskip 1em plus 0.5em minus 0.4em\relax New York, NY, USA: Association for Computing Machinery, 2020.

\bibitem{Sfakianakis2021Tracebased}
I.~Sfakianakis, E.~Kanellou, M.~Marazakis, and A.~Bilas, \emph{Trace-Based Workload Generation and Execution}.\hskip 1em plus 0.5em minus 0.4em\relax Springer International Publishing, 08 2021, pp. 37--54.

\bibitem{yildiz2024}
M.~Yildiz and A.~Baiocchi, ``Data-driven workload generation based on google data center measurements,'' in \emph{2024 IEEE 25th International Conference on High Performance Switching and Routing (HPSR)}, 2024, pp. 143--148.

\bibitem{crovella1998}
M.~Crovella, M.~Harchol-Balter, and C.~Murta, ``Task assignment in a distributed system: Improving performance by unbalancing load,'' in \emph{Proceedings of the ACM SIGMETRICS Joint International Conference on Measurement and Modeling of Computer Systems}, 1998, pp. 268--269, poster Session.

\bibitem{harcholbalter1996}
M.~Harchol-Balter, ``Network analysis without exponentiality assumptions,'' Ph.D. dissertation, University of California at Berkeley, 1996.

\bibitem{harcholbalter1996Downey}
M.~Harchol-Balter and A.~Downey, ``Exploiting process lifetime distributions for dynamic load balancing,'' in \emph{Proceedings of ACM SIGMETRICS}, Philadelphia, PA, 1996, pp. 13--24.

\bibitem{harcholbalter1999}
M.~Harchol-Balter, ``The effect of heavy-tailed job size distributions on computer system design,'' in \emph{Proceedings of ASA-IMS Conference on Applications of Heavy Tailed Distributions in Economics, Engineering and Statistics}, Washington, DC, 1999.

\bibitem{harcholbalter2003}
M.~Harchol-Balter, B.~Schroeder, N.~Bansal, and M.~Agrawal, ``Size-based scheduling to improve web performance,'' \emph{ACM Transactions on Computer Systems (TOCS)}, vol.~21, no.~2, pp. 207--233, 2003.

\bibitem{HarcholBalter2022}
M.~Harchol-Balter and Z.~Scully, ``The most common queueing theory questions asked by computer systems practitioners,'' \emph{SIGMETRICS Perform. Eval. Rev.}, vol.~49, no.~4, p. 3–7, Jun. 2022.

\bibitem{Marchal1985}
W.~Marchal, ``Numerical performance of approximate queuing formulae with application to flexible manufacturing systems,'' \emph{Annals of Operations Research}, vol.~3, no.~1, p. 141?152, 1985.

\bibitem{yildiz2025merit}
\BIBentryALTinterwordspacing
M.~Yildiz, A.~Rolich, and A.~Baiocchi, ``The merit of simple policies: Buying performance with parallelism and system architecture,'' 2025. [Online]. Available: \url{https://arxiv.org/abs/2503.16166}
\BIBentrySTDinterwordspacing

\end{thebibliography}

\end{document}